\documentclass[aps, prb, reprint]{revtex4-1}

\usepackage{fancyhdr, amsmath, amssymb, amsfonts, amsthm, bm, mathrsfs}
\usepackage{graphicx}
\usepackage{epstopdf}
\usepackage{bm}
\usepackage[citecolor=blue, colorlinks=true, urlcolor=blue, linkcolor=blue]{hyperref}

\newcommand{\ket}[1]{|#1\rangle}

\begin{document}


\title{Higgs Mode and Magnon Interactions in 2D Quantum Antiferromagnets from Raman Scattering}



\author{S. A. Weidinger}
\email[]{simon.weidinger@tum.de}

\author{W. Zwerger}
\email[]{zwerger@tum.de}

\affiliation{Physik-Department, Technische Universit\"at M\"unchen, 85747 Garching, Germany}

\begin{abstract}
We present a theory for Raman scattering on 2D quantum antiferromagnets. The microscopic Fleury-Loudon Hamiltonian is expressed in terms 
of an effective $O(3)$ - model. Well within the N\'eel ordered phase, the Raman spectrum contains a two-magnon and a two-Higgs
contribution, which are calculated diagramatically. The vertex functions for both the Higgs and magnon contributions are determined from a numerical 
solution of the corresponding Bethe-Salpeter equation.  Due to the momentum dependence of the Raman vertex in the relevant $B_{1g}+E_{2g}$ symmetry,
the contribution from the Higgs mode is strongly suppressed. Except for intermediate values of the Higgs mass,
it does not show up as separate peak in the spectrum but gives rise to a broad continuum above the dominant contribution 
from two-magnon excitations. The latter give rise to a broad, asymmetric peak at $\omega\simeq 2.44\, J$, 
which is a result of magnon-magnon interactions mediated by the Higgs mode. 
The full Raman spectrum is determined completely by the antiferromagnetic exchange coupling $J$ and a dimensionless Higgs mass. 
Experimental Raman spectra of undoped cuprates turn out to be in very good agreement with the theory only with inclusion of the Higgs 
contribution. They thus provide a clear signature of
the presence of a Higgs mode in spin one-half 2D quantum antiferromagnets. 
\end{abstract} 

%
%
\maketitle
\section{Introduction}
\label{intro}
The recent observation of a scalar Higgs particle as the final missing ingredient
of the standard model in particle physics~\cite{PhysLettB.716.1, PhysLettB.716.30} has led to a resurgence of interest in
the question whether a well defined excitation associated with an amplitude 
mode is present and observable in condensed matter systems with 
a broken continuous symmetry~\cite{PekkerVarma}.
A case in question are neutral superfluids, in which a global $U(1)$ symmetry is broken 
as the simplest example of continuous symmetry breaking. The dynamics of 
the superfluid order parameter, however,  is generically of the Gross-Pitaevskii type,
i.e. first order in time. Despite the Mexican hat structure of the effective potential,
there is thus only a Goldstone but no Higgs mode. Indeed, a necessary condition for a Higgs mode
in the non-relativistic context of condensed matter systems is an emergent Lorentz invariance.
This appears, for instance, in the superfluid phase of ultracold Bosons in an optical lattice
at integer filling, where the particle-hole symmetry close to the transition to a Mott -insulating phase
gives rise to an effectively Lorentz invariant theory~\cite{PRB.75.085106, PRL.109.010401}.
In this regime, a Higgs mode has indeed been observed from the absorbed
energy in response to shaking the optical lattice~\cite{nature11255}. The associated
effective Higgs mass vanishes in a continuous manner at the transition to a Mott insulator,  
essentially like a mirror image of the gapped particle-hole excitations in the incompressible phase.
A Higgs mode has also been observed via neutron scattering in the
antiferromagnetically ordered phase of TlCuCl$_3$ at high pressure~\cite{PRL.100.205701, naturephysics.10.373}. 
In both cases, the evolution of the Higgs mode can be followed into the
regime where the order parameter vanishes near a quantum critical point. 
Much earlier, in superconductors, the presence of an amplitude mode
has been inferred by Sooryakumar and Klein from the observation of a distinct peak in Raman 
scattering from NbSe$_2$ in a regime, where superconductivity coexists with a charge density wave~\cite{PRL.45.660}. 
In this special situation, the superconducting order parameter 
directly couples to light because the modulation of the charge density wave by the electromagnetic 
field in the $A_{1g}$ symmetry of the Raman response leads to a periodic modulation of the 
density of states at the Fermi surface and thus of the superconducting gap~\cite{PRL.47.811}.  
The amplitude mode associated with oscillations of the gap magnitude can thus 
be directly seen in Raman scattering.  This interpretation of the 
experiment has recently been confirmed within a detailed theoretical analysis by  Cea and Benfatto~\cite{PRB.90.224515}
and also in an independent experiment by M\'easson et.al.~\cite{PRB.89.060503}. \\ 

In our present paper, we analyze the possibility to observe a Higgs mode in quantum spin
models whose ground state exhibits  antiferromagnetic order in two dimensions via Raman scattering. 
This problem is of interest  for at least two different reasons:  First of all, there are experimental data on clean samples 
of undoped cuprates~\cite{PRB.78.020511, EPJST.188.131}
for which no theoretical model and understanding seems to have been given so far. Second, due to its 
relevance in the parent compounds of high temperature superconductors, quantum antiferromagnets in
two dimensions are among the most intensively studied examples of a broken continuous symmetry in condensed matter.      
In particular, the issue of a well defined Higgs excitation 
is nontrivial in two dimensions (2D). Indeed, in this case the phase space for the decay of
a Higgs mode into two Goldstone modes is very large. As a result, the Higgs spectrum at zero 
wave vector extends down to zero energy, diverging  with a universal power law $\sim 1/\omega$~\cite{PRB.49.11919, PRL.92.027203}
which has in fact been observed in neutron scattering from the N\'eel state of 
undoped La$_2$CuO$_4$~\cite{PRL.105.247001}. 
To leading order, this low energy singularity is due to two-magnon processes.
The presence of a large weight in the spectrum of the amplitude mode at low energies 
suggests that the Higgs mode is overdamped in 2D. As emphasized by Podolsky,
Auerbach and Arovas~\cite{PRB.84.174522},  this is not the case in general, however. In particular, response
functions which - unlike neutron scattering - couple to the square of the order parameter
are expected to show a clear Higgs peak even in two dimensions because for them the low frequency response
is suppressed by a factor $\omega^3$. Precisely this type of response is measured
in the lattice shaking experiment near the superfluid to Mott insulator transition performed by
Endres etal~\cite{nature11255}.  As a result, a sharp peak due to a Higgs excitation appears even in 2D,
as supported by a detailed theoretical analysis of this problem, both numerically~\cite{PRL.109.010401, PRL.110.170403, PRL.110.140401} 
and analytically~\cite{PRB.86.054508}.  In particular, the Higgs mode remains well defined even near the 
quantum critical point because both its energy and width vanish with the same power in
the distance from the critical point~\cite{PRL.109.010401, PRL.110.170403, PRL.110.140401, PRB.86.054508, PRB.89.180501}. 

Regarding the N\'eel ordered state of 2D quantum antiferromagnets,
Raman scattering seems to be ideally suited to observe the amplitude mode
since it couples to the square of the order parameter~\cite{PRB.84.174522}.
 Experimentally, however, no indications for a Higgs
mode are found~\cite{PRB.78.020511}. Instead, there is a broad, asymmetric peak
which appears to be due to a strongly renormalized two-magnon excitation. As will be shown below,
this observation can be explained both in qualitative and quantitative terms within a 
rather simple O(3) - model of the N\'eel ordered phase. 
In particular, the absence of a separate peak associated with the Higgs mode
is a consequence of the fact that the leading order contributions from the 
amplitude mode vanish because of the
nontrivial Raman vertex in the relevant  $B_{1g}$ symmetry. 
Nevertheless, the Higgs mode turns out to have a crucial effect on the detailed form of 
the Raman spectrum. In the dominant two-magnon response, it shows up 
by mediating the magnon-magnon interactions, which lead both to a downward 
shift of the peak compared to a simple spin-wave analysis~\cite{J.Phys.C.2.2012, PRB.4.922} and a
characteristic asymmetric line shape.  Moreover, the Higgs mode gives rise 
to a broad continuum above the two-magnon peak. Both features are found
in experimental spectra for La$_2$CuO$_4$~\cite{EPJST.188.131}, 
YBa$_2$Cu$_3$O$_6$~\cite{PRB.78.020511}  and also in - still unpublished - data on
Sr$_2$IrO$_4$~\cite{Gretarsson}. Using the well known
values for the antiferromagnetic exchange coupling $J$ in these systems, the  
data turn out to be in very good agreement with the theory up to frequencies
of order $6J$ and beyond. As will be shown below, the agreement between 
experiment and theory relies crucially on whether the Higgs mode contribution 
to the Raman response is included or not. Raman scattering can therefore 
indeed be used to probe the existence of a Higgs mode in 2D quantum antiferromagnets.

\section{Raman spectrum}
\label{sec:Raman spectrum}
\subsection{Effective light scattering operator}
\label{sec:Effective light scattering operator}
The differential cross section for Raman scattering is given by~\cite{RevModPhys.79.175}
\begin{equation}
\frac{d^2\sigma}{d\Omega d\omega} = \hbar r_0^2\frac{\omega_i}{\omega_f}R(i\rightarrow f)
\label{eq:CrossSection}
\end{equation}
where $r_0=e^2/mc^2$ is the Thompson radius and $R(i \rightarrow f)$ is the transition rate from the initial photon state $\ket{\mathbf{k}_i, \mathbf{e}_i}$ to the final state $\ket{\mathbf{k}_f, \mathbf{e}_f}$. The rate is determined by Fermi's golden rule~\cite{RevModPhys.79.175, PRL.65.1068}
\begin{equation}
R(i\rightarrow f)=\frac{1}{\mathcal{Z}} \sum_{I, F} e^{-\beta E_I} |M_{F, I}|^2 \delta(E_F-E_I-\hbar\omega)
\label{eq:TRate1}
\end{equation}
where $\ket{I}$, $\ket{F}$ are the initial, final state of the sample, $\hat{M}$ is the effective light scattering operator and $\omega=\omega_i-\omega_f$ is the frequency transfer. For the specific example of the Mott-insulating, antiferromagnetic phase of undoped cuprates, an appropriate microscopic 
Hamiltonian is the nearest neighbor Heisenberg model on a square lattice which arises from a superexchange mechanism \cite{PhysRev.115.2}. The associated
effective light scattering operator is then given by the Fleury-Loudon Hamiltonian in leading order of a moment 
expansion~\cite{PhysRev.166.514, PRL.65.1068, PhysLett.3.189}  Performing the common symmetry decomposition of the 
Raman scattering operator~\cite{RevModPhys.79.175}, 
it turns out that only the $B_{1g}$-mode is Raman active because the operators of the other symmetry modes commute with the Heisenberg Hamiltonian.
Restricting the interaction to this mode, the reduced effective Hamiltonian is given by
\begin{equation}
\hat{H}_\text{FL}=2 B\, P(\mathbf{e}_i, \mathbf{e}_f) \sum_j \left(\hat{\mathbf{S}}_{\mathbf{x}_j}\cdot\hat{\mathbf{S}}_{\mathbf{x}_j+\mathbf{\hat{x}}} - \hat{\mathbf{S}}_{\mathbf{x}_j}\cdot\hat{\mathbf{S}}_{\mathbf{x}_j+\mathbf{\hat{y}}}\right)
\label{eq:FL}
\end{equation}
with $P(\mathbf{e}_i, \mathbf{e}_f)=e_i^x e_f^x - e_i^y e_f^y$ and $B\sim t^2/(U-\hbar\omega_i)$. 
As has been shown by Shastry and Shraiman\cite{PRL.65.1068}, 
there is also a response in other symmetry modes associated with ring exchanges on a plaquette or the fluctuations of a chiral 
spin operator
\footnote{Regarding the latter, a contribution from a spin-chirality term $\sim\mathbf{S}_i\cdot(\mathbf{S}_j\wedge\mathbf{S}_k)$
is not present in Raman scattering from cuprates, where the magnetic order arises from electrons on a square lattice with 
nearest neighbor hopping. It may appear, however, in more  complex systems like a Kagom\'e lattice where it 
may be relevant to detect spin-liquid phases, see~\cite{PRB.81.024414}. }.   
They require an extension of the microscopic Hamiltonian beyond the Heisenberg model with nearest neighbor exchange only and thus 
lead to corrections to the simple form of the Fleury-Loudon Hamiltonian~\eqref{eq:FL}.
In our present work, we do not consider such extensions. In fact, as will be shown below, already the leading order 
Hamiltonian~\eqref{eq:FL}, which applies to 2D quantum antiferromagnets with a well defined N\'eel order, 
properly accounts for the essential features of the Raman spectrum. 
Apart from microscopic prefactors, 
the transition rate in Eq.~\eqref{eq:TRate1} can then be written as the Fourier transform of the van Hove function $S(t)$ of $\hat{H}_\mathrm{FL}$
\begin{align}
R(i\rightarrow f)&=\frac{e^2}{\hbar^3 c^2}g(k_i)g(k_f)\int\limits_{-\infty}^{\infty}dt \text{ } e^{i\omega t} \langle \hat{H}_\text{FL}(t) \hat{H}_\text{FL}(0)\rangle_T\notag\\
&=\frac{e^2}{\hbar^3 c^2}g(k_i)g(k_f)S(\omega)
\label{eq:TRate2}
\end{align}
The resulting Raman response function $S(\omega)$ is connected to an associated spectral function $\chi''_\text{Raman}(\omega)$
by the fluctuation-dissipation theorem. At zero temperature, this
takes the simple form
\begin{equation}
S(\omega)=2\hbar\chi''_\text{Raman}(\omega)\theta(\omega)\equiv 2\hbar\text{ Im}\chi_\text{Raman}(\omega+i0)\theta(\omega)
\end{equation}
In practice, the spectral function will be 
determined from the imaginary time ordered correlation function 
\begin{equation}
\chi_\text{Raman}(\tau)=\frac{1}{\hbar}\left\langle T_\tau \left[\hat{H}_\text{FL}(\tau)\hat{H}_\text{FL}(0)\right]\right\rangle
\end{equation}
by analytic continuation to real frequencies. 

\subsection{Field theoretic formalism}
\label{sec:Field theoretic formalism}
In order to disentangle the contributions from magnons and the Higgs mode to the Raman spectrum, it is both
necessary and convenient to replace the spin-operators of the microscopic Heisenberg model by a coarse grained description
in which the relevant excitations appear more directly. Such an effective description is provided by the $O(3)$ 
nonlinear $\sigma$-model (NL$\sigma$M), which is the continuum field theory of the antiferromagnetic Heisenberg model as used by Chakravarty, Halperin and Nelson~\cite{PRB.39.2344}. 
The associated effective action can be written in an apparently relativistic invariant form
\begin{equation}
S\left[\mathbf{n}\right]=\frac{1}{2g}\int_\Lambda d^3x~\left(\partial_\mu\mathbf{n}\right)^2 \text{   ,  } g=\frac{\hbar c_s}{\rho_s}
\label{eq:NLM}
\end{equation}
with a three-vector $(x^\mu)=(c_s\tau, \mathbf{x})$ and spin wave velocity $c_s$. Apart from $c_s$ and the momentum cutoff $\Lambda$,
the dimensionless coupling constant $\tilde{g}=g\Lambda$ of the nonlinear $\sigma$-model contains the spin stiffness $\rho_s$, which is of the order of 
the exchange coupling constant of the underlying Heisenberg model. The slowly varying order parameter field $\mathbf{n}(x)$ is 
a three component field which describes
deviations from perfect N\'eel order. It is normalized by $\mathbf{n}^2=1$.
On the microscopic level, the physical meaning of the field $\mathbf{n}$ becomes evident by the mapping 
\begin{equation}
\frac{\hat{\mathbf{S}}(\mathbf{x}_j)}{\hbar S} \rightarrow (-1)^{\mathrm{sgn}( \mathbf{x}_j)} \mathbf{n}(\mathbf{x}_j) \sqrt{1-a^{4}\mathbf{L}^2(\mathbf{x}_j)}+a^2 \mathbf{L}(\mathbf{x}_j)
 \label{eq:Haldane}
\end{equation}
between spin operators in the semiclassical limit $S\gg 1$ and bosonic fields originally due to Haldane \cite{PRL.50.1153, PhysLettA.93.464}.
Here $a$ is the lattice constant and the angular momentum field $\mathbf{L}$ is constrained by 
$\mathbf{n}\cdot\mathbf{L}=0$, $a^{4}\mathbf{L}^2\ll1$.  Using Eq.~\eqref{eq:Haldane} to express the spin operators in the 
Fleury-Loudon Hamiltonian~\eqref{eq:FL} in terms of the order parameter field, it turns out that the Raman spectrum 
requires to determine the connected correlation function 
of the following operator in the NL$\sigma$M
\begin{align}
O(\tau) &= 2B\,\hbar^2S^2 P(\mathbf{e}_i, \mathbf{e}_f) \int d^2x \text{ }\sigma_{ij}^z \partial_i \mathbf{n}(\mathbf{x}, \tau)\cdot\partial_j \mathbf{n}(\mathbf{x}, \tau)\notag
\\
&=4B\, \hbar^2 S^2 P(\mathbf{e}_i, \mathbf{e}_f) \int\frac{d^2 k}{(2\pi)^2} \gamma(\mathbf{k}) \left|\mathbf{n}(\mathbf{k}, \tau)\right|^2
\label{eq:RamanOxy}
\end{align}
Here $\gamma(\mathbf{k})=\tfrac{1}{2}(k_x^2-k_y^2)$ is the continuum form of the $B_{1g}$-symmetry factor. Unfortunately,
the combination of nontrivial vertices in the NL$\sigma$M which appear beyond a simple Gaussian approximation
due to the constraint $\mathbf{n}^2=1$ and the complicated form of the operator~\eqref{eq:RamanOxy} 
make it very hard to calculate the Raman spectrum directly within the NL$\sigma$M. Moreover, it is 
rather difficult to disentangle the contributions which arise from magnons or the Higgs excitation
in the standard decomposition $\mathbf{n}=(\bm{\pi}, \sqrt{1-\bm{\pi}^2})$ which only involves the two-component field $\bm{\pi}$
associated with the magnons. It is therefore more convenient to use 
a soft-spin description of the order parameter within a linear $O(3)$ $\sigma$-model, whose effective action 
\begin{align}
S\left[\bm{\Phi}\right] = \frac{1}{2g} \int_\Lambda d^{3} x \left[\left(\partial_\mu\bm{\Phi}\right)^2+\frac{m_0^2}{12}\left(\left|\bm{\Phi}\right|^2-3\right)^2\right]
\label{eq:linearON}
\end{align}
contains the bare mass $m_0$ of the amplitude mode as an additional parameter. It determines the stiffness
for longitudinal fluctuations in the N\'eel ordered phase which arise from the fact that for quantum spins
the operator $\hat{S}^z_{\mathbf{Q}_{\rm AF}}$ which determines the long range antiferromagnetic order at the 
wave vector $\mathbf{Q}_{\rm AF}$ has finite fluctuations along the $z$-direction
\footnote{For a discussion of these fluctuations on a microscopic basis and in the context of neutron scattering, see~\cite{PRB.72.224511}.}  
In general, the coupling constant $g$ and $m_0$ are independent parameters. It is only 
in the scaling regime close to the quantum critical point $\tilde{g}=\tilde{g}_c$ for the 
loss of N\'eel order, where the value of the bare Higgs mass is irrelevant and where the $O(3)$ $\sigma$-model 
is completely equivalent to the NL$\sigma$M~\cite{PRB.86.054508}.
In fact, as will be shown below, the dominant features of the Raman spectrum are not very sensitive to the value of the Higgs mass $m_0$ even in the opposite regime $\tilde{g}\ll \tilde{g}_c$ of well defined N\'eel order. They are determined essentially by the spin stiffness or the equivalent Heisenberg exchange energy $J$, which is known rather accurately 
from neutron scattering data. 
An intuitive argument for why it is possible to replace the fixed length NL$\sigma$M 
with a soft-spin model relies on viewing the linear $\sigma$-model as an effective low-energy theory of the NL$\sigma$M, 
where short wave-length fluctuations have been integrated out~\cite{Sachdev}.  The Ne\'el-field $\mathbf{n}(\mathbf{x})$ is
 therefore averaged over some domain in real space and the resulting averaged field 
$\tilde{\bm{\Phi}}$ is no longer constrained to have a fixed length. Instead, it is subject to a potential $V(\tilde{\bm{\Phi}})$, 
which exhibits a minimum at a finite vacuum expectation value (VEV), which has been chosen to be $\langle\left|\bm{\Phi}\right|\rangle=\sqrt{3}$ in Eq.~\eqref{eq:linearON}. The magnitude of the fluctuations around the minimum is controlled by the 
dimensionless mass parameter $\tilde{m}_0=m_0/\Lambda$, which is expected to be large compared to one in the 
semiclassical limit $S\gg 1$, where fluctuations of the order parameter magnitude vanish.
The dynamics of the averaged field is fixed by the condition that the magnon dispersion $\omega_k=c_s|\mathbf{k}|$ is
linear in momentum, giving rise to an effectively Lorentz invariant action. 

In order to calculate the Raman spectrum within the linear $\sigma$-model~\eqref{eq:linearON}, 
we use the standard parametrization for the field $\bm{\Phi}$ in the symmetry broken phase:
\begin{equation}
\bm{\Phi}=\left(\bm{\pi}, r\sqrt{3} + \sigma\right)\, .
\label{eq:para}
\end{equation}
The parameter $r=r(g, \Lambda)$ incorporates the renormalization of the VEV by quantum fluctuations and has to be determined from the condition $\langle\sigma\rangle=0$. The components of $\bm{\pi}$ are the two massless antiferromagnetic magnons, while $\sigma$ denotes the massive amplitude or 
Higgs mode. Expressed in terms of the relativistic three-momentum $k=(\omega/c_s,\mathbf{k})$, the free propagators for the fields are
\begin{align}
&G_{\pi_i\pi_j}^0(k)=\langle\pi_i(k)\pi_j(-k)\rangle=\delta_{ij} G_{\pi\pi}^0(k)=\delta_{ij}\frac{g}{k^2}\label{eq:MPropLO}\\
&G_{\sigma\sigma}^0(k)=\langle\sigma(k)\sigma(-k)\rangle=\frac{g}{k^2+m_0^2}\label{eq:HPropLO}
\end{align}
These propagators will change with increasing strength $g$ of the quantum fluctations. In particular, the exact propagator
 \begin{equation}
G_{\sigma\sigma}(k)=\frac{g}{k^2+m_0^2}+\frac{g^2}{\left(k^2+m_0^2\right)^2}\frac{m_0^4}{24 |k|}+\ldots
\label{eq:exactGsigma}
\end{equation} 
of the Higgs mode acquires a contribution $\sim 1/|k|$ due to the decay into Goldstone modes. 
Instead of a simple pole, the Higgs propogator therefore exhibits a branch cut which starts at the threshold $\omega_k=c_s|\mathbf{k}|$ for the excitation 
of magnons~\cite{PRL.92.027203}. This low energy singularity 
first appears at order $g^2$ and is expected to remain valid to all orders, 
i.e. $G_{\sigma\sigma}(k\to 0)$ is proportional to $1/|k|$ at any finite value of $g$.
Similarly, by the Goldstone theorem, the magnons are always massless. The structure $G_{\pi\pi}(k) =Zg/k^2$ 
therefore remains exact with a finite, renormalized 
coupling constant $Z g$ throughout the N\'eel ordered phase\footnote{For a recent discussion of infrared singularities in the context of neutral fermionic superfluids see \cite{PRB.88.144508}.}. 
Inserting Eq.~\eqref{eq:para} into Eq.~\eqref{eq:RamanOxy} we obtain 
\begin{align}
&O(\tau)=O_\text{M}(\tau)+O_\text{H}(\tau)\label{eq:RamanO}\\
&O_\text{M}(\tau)\propto \int d^2x \left[\left(\partial_x\bm{\pi}\right)^2 - \left(\partial_y\bm{\pi}\right)^2 \right]\\
&O_\text{H}(\tau) \propto \int d^2x \left[ \left(\partial_x\sigma\right)^2 - \left(\partial_y\sigma\right)^2\right]
\end{align}
The operator to which light couples in the Raman response therefore involves the square of gradients of  
the magnon and Higgs field. As a result, the response always involves at least two magnon or Higgs excitations.   
Taken together, the full Raman susceptibility is the sum of three different contributions
\begin{equation}
\chi_\text{Raman}(\omega)=\chi_{\text{M}}(\omega) + \chi_{\text{H}}(\omega) + \chi_\text{Int}(\omega)
\label{eq:RamanSusceptSum}
\end{equation}
which involve a two-magnon susceptibility $\chi_\text{M}$ and a two-Higgs susceptibility $\chi_\text{H}$. They are determined by the correlation functions of two Magnon and two Higgs operators $O_\text{M}$ and $O_{H}$, respectively. In addition, an interference susceptibility $\chi_\text{Int}$ arises which consists of the two mixed correlation functions. A similar structure has been found by Canali and Girvin \cite{PRB.45.7127}, 
who have calculated the Raman spectrum of undoped cuprates by means of a Dyson-Maleev transformation
of the underlying antiferromagnetic Heisenberg model. Compared to our present formulation,
this approach does not allow to separate the magnon and Higgs contributions to the spectrum, with the latter appearing 
as part of a four-magnon term. Moreover, the magnon-magnon interaction is 
replaced by an instantaneous one, thus missing the retardation effects associated with the Higgs-mediated 
interaction discussed below which is treated properly by our numerical solution of the 
Bethe-Salpeter equation.\\

\begin{figure}
\centering
\includegraphics[scale=0.6]{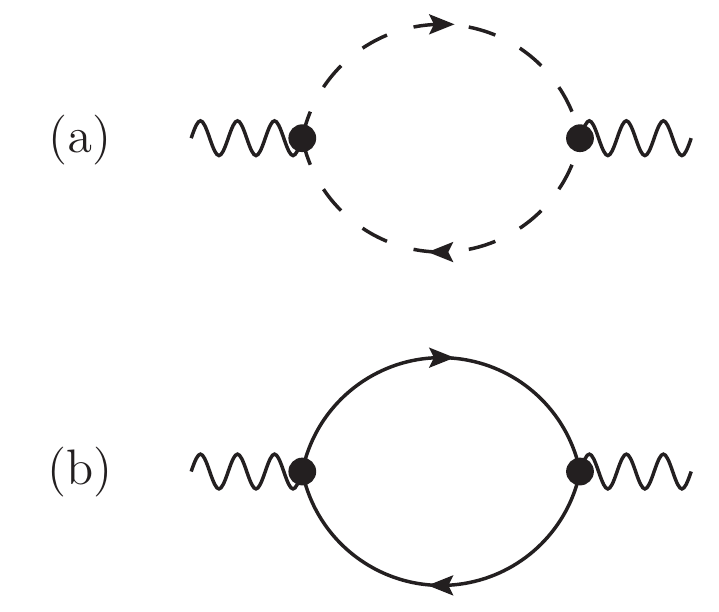}
\caption{Leading order diagram for the two-magnon (a) and the two-Higgs susceptibility (b). The dashed line propagators are
those of the bare magnons in Eq.~\eqref{eq:MPropLO}, the full lines denote the Higgs propagator in Eq.~\eqref{eq:HPropLO}. }
\label{fig:Bubbles}
\end{figure}

 To leading order in the coupling $g$, only the two-magnon and two-Higgs susceptibilities are nonzero while
 the interference susceptibility contributes only at order $g^3$. The corresponding Feynman diagrams in  FIG.~\ref{fig:Bubbles} 
 are just a two-magnon, respectively a two-Higgs bubble dressed with two Raman symmetry factors $\gamma(\mathbf{k})$.
Dropping the polarization factor $P(\mathbf{e}_i, \mathbf{e}_f)$, the two-magnon response to lowest order in $g$ is given by
\begin{equation}
\chi_{\text{M}}''(\omega)=\frac{g^2B^2\hbar^3S^4}{12c_s}\omega^3 \theta(2 c_s \Lambda-\omega)+O(g^3)\, .
\label{eq:2MagnonLeading}
\end{equation}
The spectrum is a pure power law  $\sim\omega^3$ with a sharp cutoff at twice the zone-boundary energy of a magnon. If we assume, that the Heisenberg model underlying the continuum theory contains only a nearest-neighbor coupling $J$, this corresponds to $2c_s\Lambda=2\pi\hbar J$. 

Regarding the contribution from longitudinal fluctuations of the order parameter, the spectrum to leading order in $g$ 
is determined by the diagram in FIG.~\ref{fig:Bubbles} b, which gives
\begin{align}
\chi_{\text{H}}''(\omega)=&\frac{g^2 B^2\hbar^3S^4}{12c_s\omega}\left[\left(\frac{\omega}{2}\right)^2-c_s^2 m_0^2\right]^2 \theta(\omega-2c_sm_0)\notag\\
&\times\theta\left(2c_s\sqrt{\Lambda^2+m_0^2}-\omega\right) + O(g^3)\, . \label{eq:2HiggsLO}
\end{align}
It shows the expected threshold at twice the Higgs mass and peaks at the maximum energy of two Higgs excitations. 
Note that in contrast to neutron scattering, which couples linearly to the order parameter, the Raman spectrum does not contain the 
longitudinal propagator $G_{\sigma\sigma}$ directly.  
 As a result, no sharp peak due to the Higgs mode appears even at leading order. In the following section, we will 
 show that the sharp features in both the two-magnon and the two-Higgs response, which depend quite sensitively on the precise form of the momentum cutoff and the specific value of the Higgs mass $m_0$, are completely
eliminated in a calculation which sums up the next-to-leading order diagrams in a consistent fashion.     
 

\subsection{Beyond leading order and Bethe-Salpeter equation}
\label{sec:Beyond leading order and Bethe-Salpeter equation}

\begin{figure}
\centering
\includegraphics[scale=0.35]{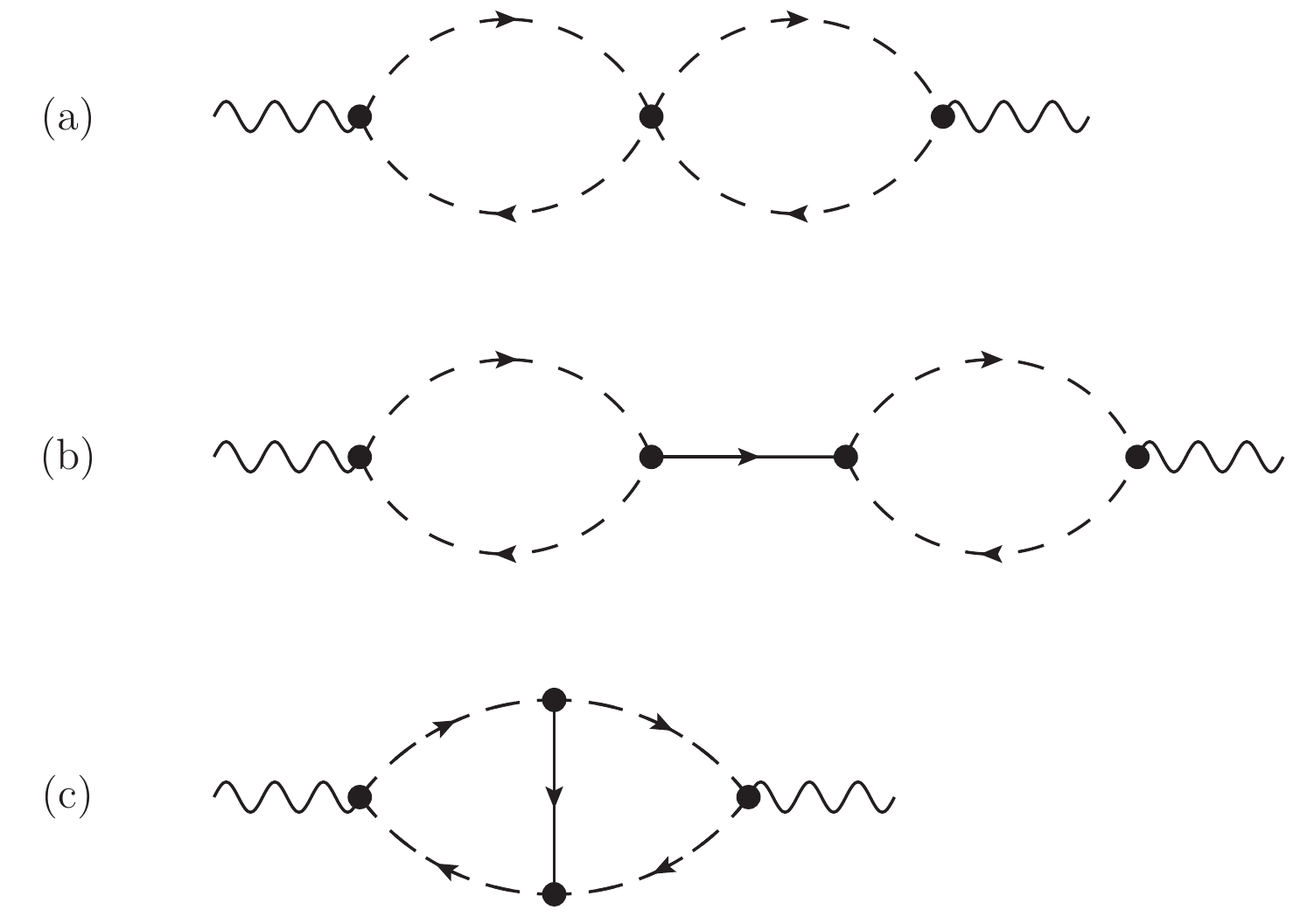}
\caption{Next-to-leading order diagrams apart from self-energy insertions for the 2-magnon susceptibility $\chi_{\text{M}}$}
\label{fig:2MagnonNLO}
\end{figure}

\begin{figure}
\centering
\includegraphics[scale=0.35]{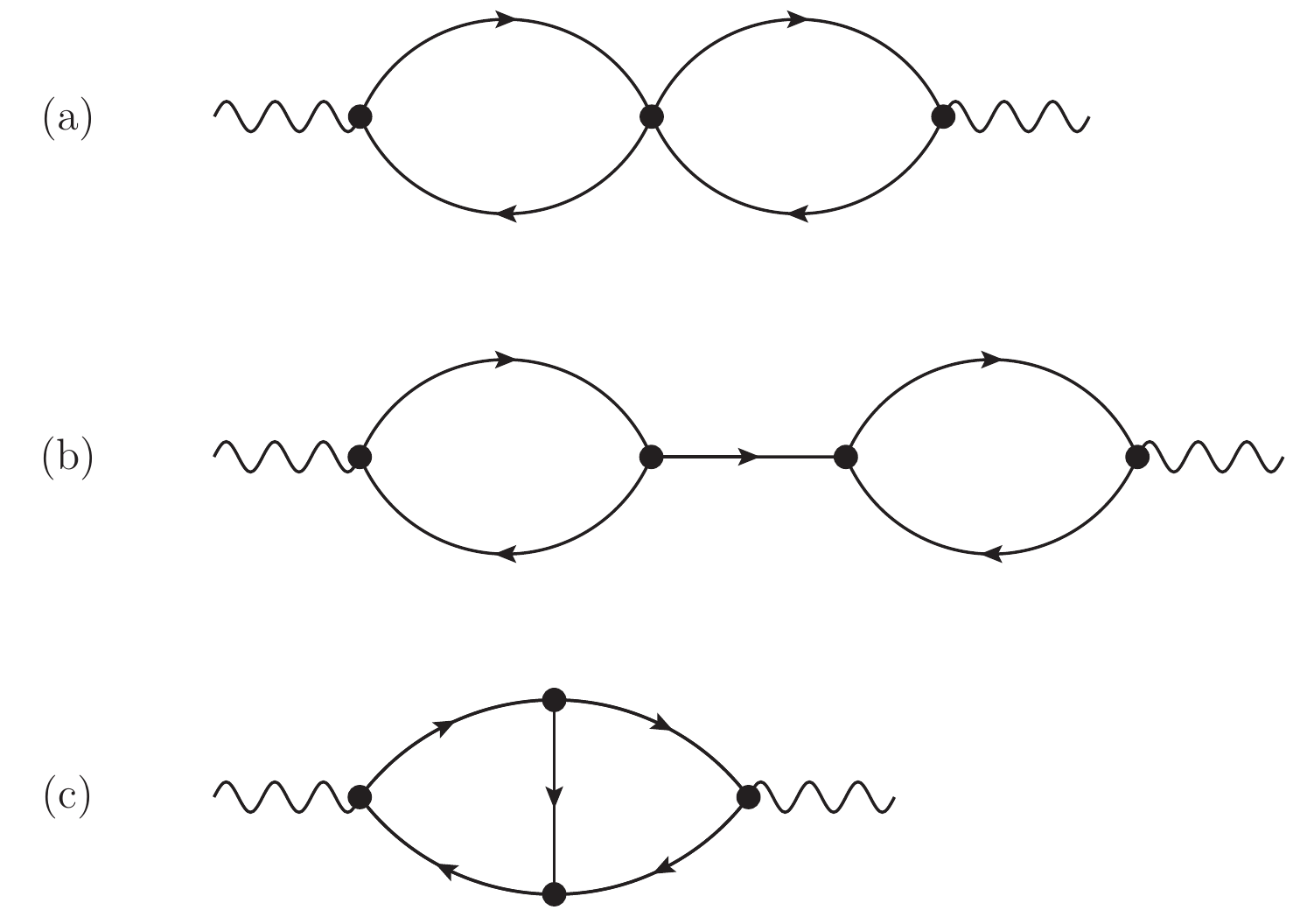}
\caption{Next-to-leading order diagrams apart from self-energy insertions for the 2-Higgs susceptibility $\chi_{\text{H}}$}
\label{fig:2HiggsNLO}
\end{figure}

\begin{figure}
\centering
\includegraphics[scale=0.35]{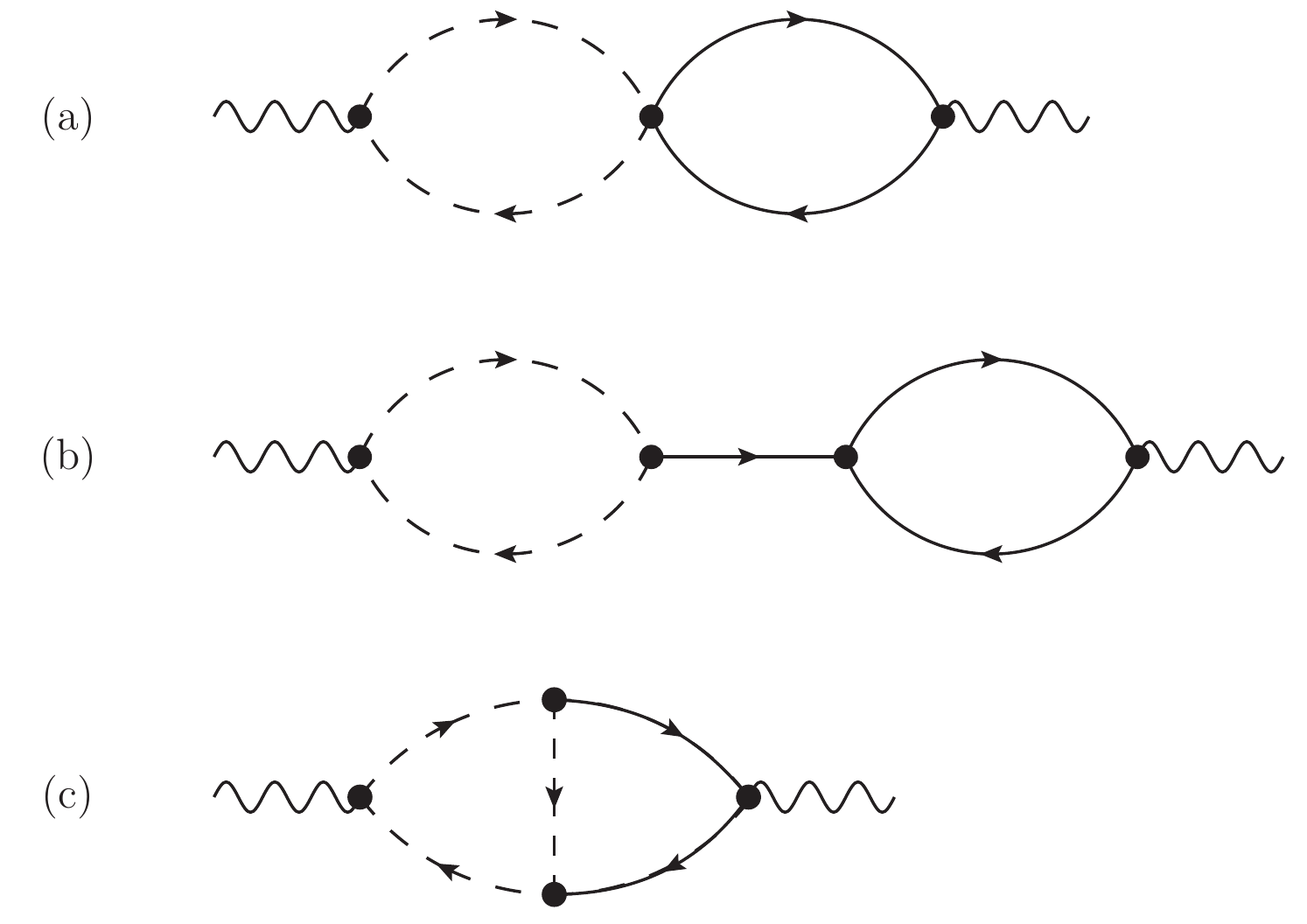}
\caption{Next-to-leading order diagrams for the interference susceptibility $\chi_{\text{Int}}$.}
\label{fig:InterferenceNLO}
\end{figure}

The next-to-leading order diagrams, apart from self-energy insertions,  for the two-magnon, two-Higgs and interference susceptibility are shown in figures FIG.~\ref{fig:2MagnonNLO}, FIG.~\ref{fig:2HiggsNLO} and FIG.~\ref{fig:InterferenceNLO} respectively. To see whether these diagrams may lead to sharp 
features associated with the Higgs mode, we start with a qualitative discussion based on symmetries alone. One immediately notices, that the loop integrals in FIG.~\ref{fig:2MagnonNLO}(a) decouple at vanishing external wave vector, where $q=(\omega,0)$. 
These diagrams therefore give no contribution to the Raman response 
because of the antisymmetry of the $B_{1g}$-symmetry factor $\tilde{\gamma}(\mathbf{k})$ under the exchange $k_x\leftrightarrow k_y$.

Especially the vanishing of diagram FIG.~\ref{fig:2MagnonNLO}(b) is important to notice, as it would lead to a peak at the Higgs mass 
due to the pole in the Higgs propagator at least in leading order in $g$. 
The only non-vanishing diagram for the two-magnon susceptibility is FIG.~\ref{fig:2MagnonNLO}(c), in which the Higgs line is independent of the frequency transfer.  As a result, no sharp peak at the Higgs mass arises. For the two-Higgs and the interference susceptibilities the situation is quite similar. Indeed, the diagrams (a) and (b) in FIG.~\ref{fig:2HiggsNLO} and FIG.~\ref{fig:InterferenceNLO} vanish due to the antisymmetry of the $B_{1g}$-symmetry factor under the exchange $k_x\leftrightarrow k_y$. 
Concerning the interference contribution, it turns out, that up to order $g^3$ it gives a negligible contribution to the full Raman response.
Thus, in the following, we only consider the two-magnon and two-Higgs contributions.

\begin{figure}
\includegraphics[scale=0.75]{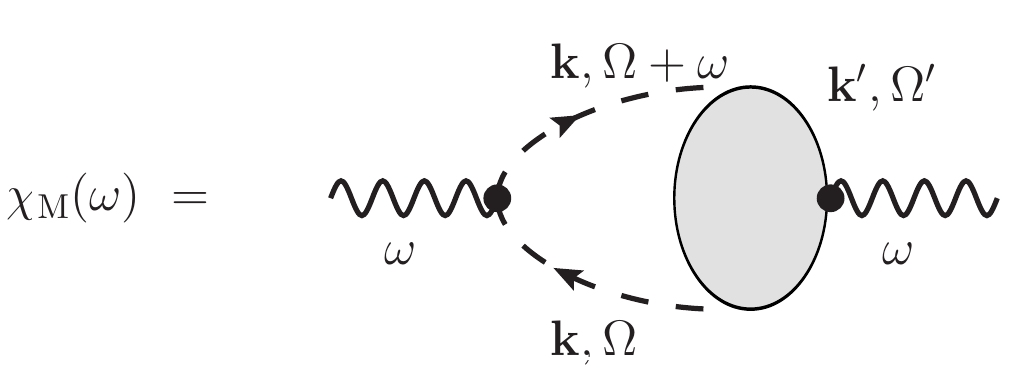}
\caption{Diagrammatic series for the 2-magnon susceptibility. Magnon lines have to be interpreted as full propagators. The shaded area represents the full 2-magnon Raman vertex $\Gamma_\pi$.}
\label{fig:MagnonBS}
\end{figure}

\begin{figure}
\includegraphics[scale=0.47]{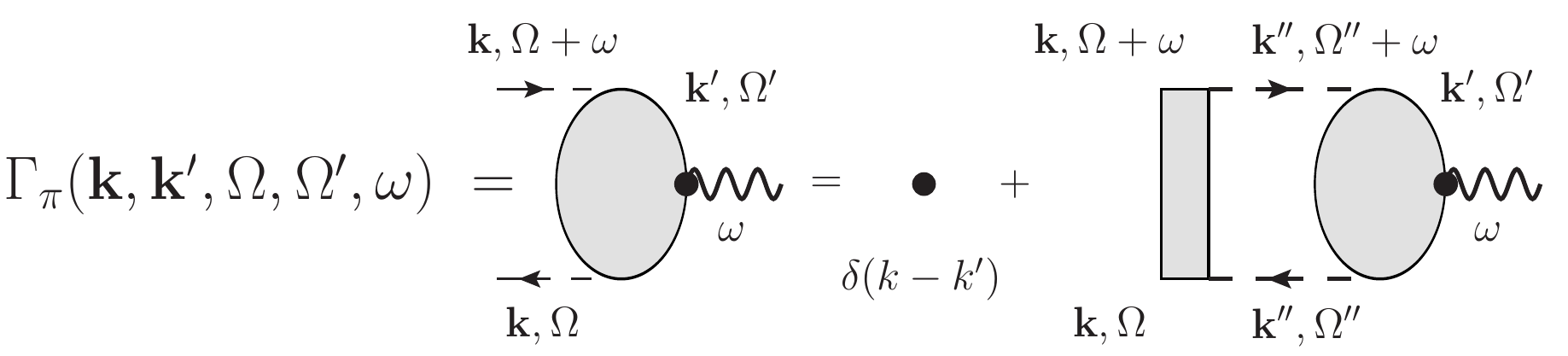}
\caption{Diagrammatic series for the full 2-magnon Raman vertex, giving the Bethe-Salpeter equation \eqref{eq:BSMagnonVertexF}.}
\label{fig:MagnonVertex}
\end{figure}

In order to determine the Raman spectrum in quantitative terms, we now derive a formally exact equation for the two-magnon and the two-Higgs response. 
Based on the diagrammatic expansion shown in FIG.~\ref{fig:MagnonBS}, the two-magnon contribution $\chi_{\text{M}}$ can be written in terms of the exact two-magnon Raman vertex function $\Gamma_\pi(\mathbf{k}, \mathbf{k}', \Omega, \Omega', \omega)$ in the following form

\begin{align}
\chi_{\text{M}}(\omega)=2\int\frac{d^3 k}{(2\pi)^3}&\int\frac{d^3 k'}{(2\pi)^3} \tilde{\gamma}(\mathbf{k})\tilde{\gamma}(\mathbf{k}') G_{\pi\pi}(\mathbf{k}, \Omega+\omega)\notag\\
 &\times G_{\pi\pi}(\mathbf{k}, \Omega) \Gamma_{\pi}(\mathbf{k}, \mathbf{k}', \Omega, \Omega', \omega)\, .
\label{eq:BS2MagnonF}
\end{align}
The vertex function obeys a Bethe-Salpeter equation 
\begin{align}
\Gamma_\pi(\mathbf{k}, \mathbf{k}&', \Omega, \Omega', \omega)=2(2\pi)^3 \delta(k-k') + \int\frac{d^3k''}{(2\pi)^3} \mathcal{V}_\pi(k, k'', \omega)\notag\\ 
&\times G_{\pi\pi}(\mathbf{k}'', \Omega''+\omega)G_{\pi\pi}(\mathbf{k}'', \Omega'') \Gamma_\pi(\mathbf{k}'', \mathbf{k}', \Omega'', \Omega', \omega)
\label{eq:BSMagnonVertexF}
\end{align}
which is depicted diagrammatically in FIG.~\ref{fig:MagnonVertex}.
To determine the Raman spectrum, we do not need the full vertex function. Instead, it is sufficient to know the reduced one, which is defined as 
\begin{equation}
\Gamma_\pi(\mathbf{k}, \Omega, \omega)\equiv \int\frac{d^3k'}{(2\pi)^3}\tilde{\gamma}(\mathbf{k}')\Gamma_\pi(\mathbf{k}, \mathbf{k}', \Omega, \Omega', \omega)
\label{eq:2MagnonVert}
\end{equation}
Using this definition, the equations for the two-magnon susceptibility, Eq.'s~\eqref{eq:BS2MagnonF} and \eqref{eq:BSMagnonVertexF}, can be recast 
in the form
\begin{align}
\chi_{\text{M}}(\omega)=2\int\frac{d^3 k}{(2\pi)^3} &\tilde{\gamma}(\mathbf{k})G_{\pi\pi}(\mathbf{k}, \Omega+\omega)\notag\\ 
&\times G_{\pi\pi}(\mathbf{k}, \Omega) \Gamma_{\pi}(\mathbf{k},  \Omega, \omega)\label{eq:BS2MagnonR}
\end{align}
\begin{align}
\Gamma_\pi(\mathbf{k}, \Omega, \omega)= 2\tilde{\gamma}&(\mathbf{k}) + \int\frac{d^3k''}{(2\pi)^3} \mathcal{V}_\pi(k, k'', \omega) G_{\pi\pi}(\mathbf{k}'',\Omega'')\notag\\
&\times G_{\pi\pi}(\mathbf{k}'', \Omega''+\omega) \Gamma_\pi(\mathbf{k}'', \Omega'', \omega)\label{eq:BSMagnonVertexR}
\end{align}
Up to this point, there are no approximations involved.  A solution of Eq. \eqref{eq:BSMagnonVertexR}, 
however requires knowledge of both the exact magnon propagator 
\begin{align}
G_{\pi\pi}(p)=\frac{g}{p^2-g\Sigma_{\pi\pi}(p)}
\end{align}
including all self-energy corrections via the exact magnon self-energy $\Sigma_{\pi\pi}$   
and the full irreducible two-magnon interaction $\mathcal{V}_\pi(k, k'', \omega)$.


For an explicit and quantitative solution,
we truncate the irreducible magnon interaction at lowest order in $g$, which amounts to a ladder approximation. 
Properly taking into account combinatorial factors we find
\begin{equation}
\mathcal{V}_{\pi}(k, k'', \omega)=2\cdot 4 \cdot \frac{1}{2} \left(\frac{m_0^2}{2\sqrt{3}g}\right)^2 G_{\sigma\sigma}^0(k-k'') 
\label{eq:irredIntMagnonLadder}
\end{equation}
Thus, to lowest order, the interaction of magnons which is relevant for the $B_{1g}$ symmetry is mediated by the Higgs mode. Inserting Eq.~\eqref{eq:irredIntMagnonLadder} into Eq.~\eqref{eq:BSMagnonVertexR}, we obtain the ladder Bethe-Salpeter equation, which corresponds to a consistent next-to-leading order approximation in the irreducible magnon-magnon interaction.  For the magnon propagator, the zeroth order 
result $G_{\pi\pi}^0$ captures the exact low energy behavior $\sim g/|k|^2$ due to the Goldstone theorem apart from a finite 
renormalization factor $g\to Zg$. 
Regarding the Higgs-propagator, the leading order result contains a simple pole at the bare Higgs mass $m_0$ but misses 
the $1/|k|$-singularity due to the decay of the Higgs into two Goldstone modes, see Eq. \eqref{eq:exactGsigma}. 
The final form of the Bethe-Salpeter equation in ladder approximation is given by
\begin{align}
\Gamma_\pi&^\text{Ladder}(\mathbf{k}, \Omega, \omega)= 2\tilde{\gamma}(\mathbf{k}) +\frac{m_0^4}{3g^2} \int\frac{d^3k''}{(2\pi)^3} G^0_{\sigma\sigma}(k-k'')\notag\\ 
&\times G^0_{\pi\pi}(\mathbf{k}'', \Omega'')G^0_{\pi\pi}(\mathbf{k}'', \Omega''+\omega)\Gamma_\pi^\text{Ladder}(\mathbf{k}'', \Omega'', \omega)\, .
\label{eq:2MagnonLadderV}
\end{align}
This is still too complicated to be solved analytically, however it is amenable to an efficient numerical solution. Before, doing so, 
however,  a similar equation will be derived for the two-Higgs susceptibility. \\

\begin{figure}
\includegraphics[scale=0.6]{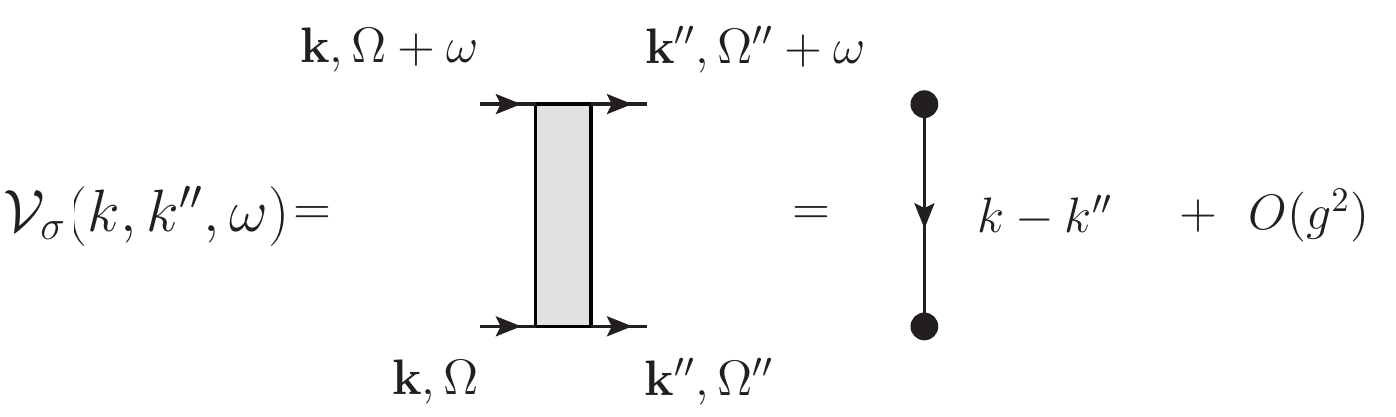}
\caption{Diagrammatic representation of the irreducible 2-Higgs interaction vertex.}
\label{fig:HiggsInt}
\end{figure}

The way to derive the integral equations for the two-Higgs susceptibility is completely analogous to the case of the two-Magnon susceptibility, one just needs to replace the magnon propagators by Higgs propagators. Hence we will just write down the resulting equations, containing already the reduced two-Higgs Raman vertex function $\Gamma_\sigma(\mathbf{k}, \Omega, \omega)$:
\begin{align}
\chi_{\text{H}}(\omega)=2\int\frac{d^3 k}{(2\pi)^3} \tilde{\gamma}(\mathbf{k})&G_{\sigma\sigma}(\mathbf{k}, \Omega+\omega) G_{\sigma\sigma}(\mathbf{k}, \Omega)\notag\\
&\times \Gamma_{\sigma}(\mathbf{k},  \Omega, \omega)
\end{align}
\begin{align}
\Gamma_\sigma(\mathbf{k}, \Omega, \omega)= 2\tilde{\gamma}&(\mathbf{k}) + \int\frac{d^3k''}{(2\pi)^3} \mathcal{V}_\sigma(k, k'', \omega)G_{\sigma\sigma}(\mathbf{k}'', \Omega'')\notag\\ &\times G_{\sigma\sigma}(\mathbf{k}'', \Omega''+\omega)\Gamma_\sigma(\mathbf{k}'', \Omega'', \omega)
\label{eq:BSHiggsVertexR}
\end{align}
The leading diagrams for the full irreducible two-Higgs interaction $\mathcal{V}_{\sigma}(k, k'', \omega)$ are depicted in FIG.~\ref{fig:HiggsInt}. Truncating the irreducible two-Higgs interaction again at leading order in $g$, we obtain
\begin{align}
\mathcal{V}_{\sigma}(k, k'', \omega)=2\cdot(2\cdot 3)^2  \cdot \frac{1}{2} \left(\frac{m_0^2}{2\sqrt{3}g}\right)^2 G^0_{\sigma\sigma}(k-k'') 
\label{eq:irredIntHiggsLadder}
\end{align}
Note that this has precisely the same form as the irreducible two-magnon interaction, differing only in the combinatorial and coupling prefactors.
Inserting eq.~\eqref{eq:irredIntHiggsLadder} into Eq.~\eqref{eq:BSHiggsVertexR} we obtain the Bethe-Salpeter equation for the reduced two-Higgs Raman vertex function
\begin{align}
\Gamma_\sigma&^\text{Ladder}(\mathbf{k}, \Omega, \omega)= 2\tilde{\gamma}(\mathbf{k}) +\frac{3m_0^4}{g^2} \int\frac{d^3k''}{(2\pi)^3} G^0_{\sigma\sigma}(k-k'')\notag\\
&\times G^0_{\sigma\sigma}(\mathbf{k}'', \Omega'')G^0_{\sigma\sigma}(\mathbf{k}'', \Omega''+\omega)\Gamma_\sigma^\text{Ladder}(\mathbf{k}'', \Omega'', \omega)
\label{eq:2HiggsLadderV}
\end{align}
which again needs to be solved numerically. 

\subsection{Numerical solution of the Bethe-Salpeter equation and Raman spectra}
\label{sec:Numerical solution of the Bethe-Salpeter equation and Raman spectra}

\begin{figure}
\includegraphics[width=\linewidth, height=200pt]{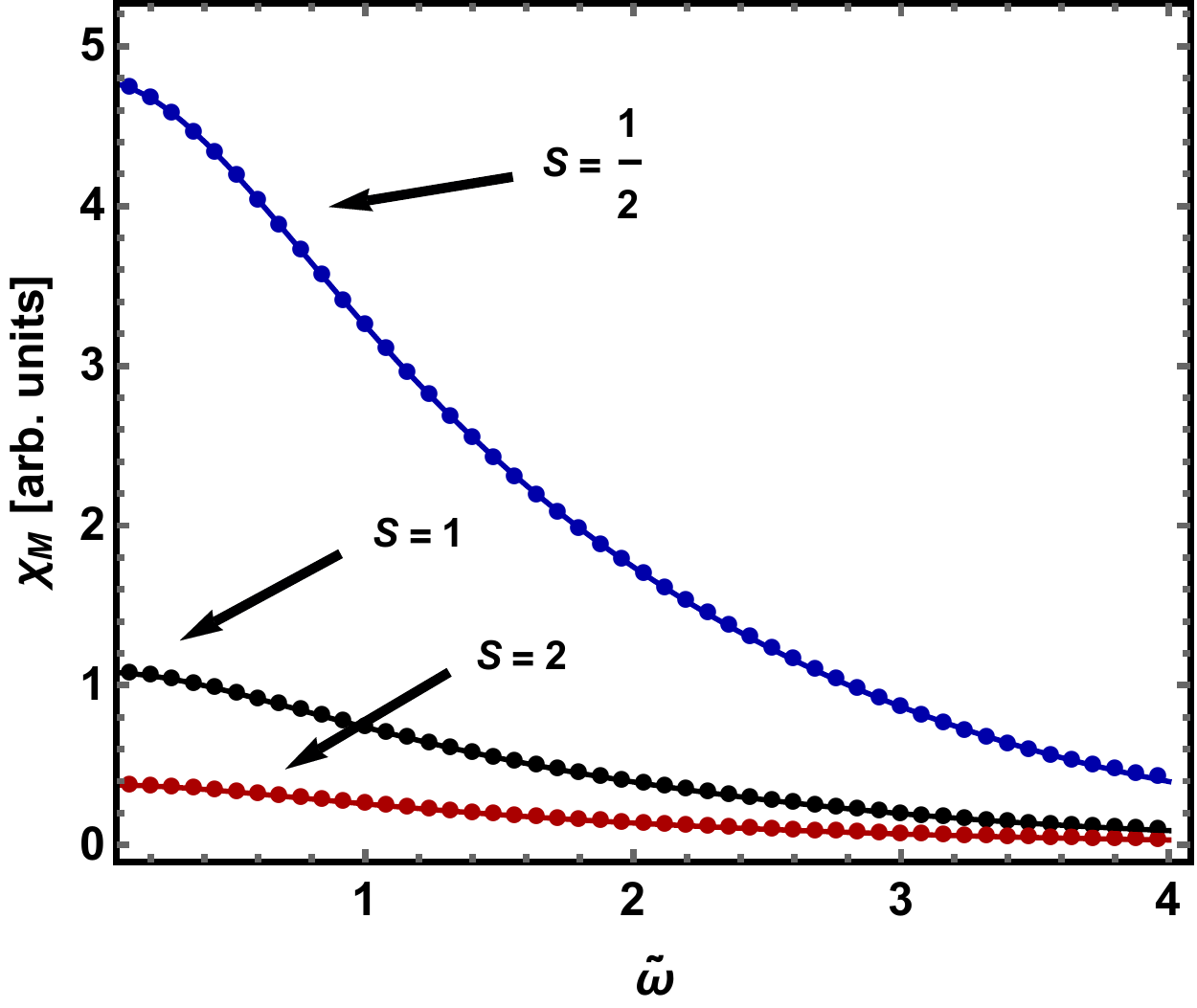}
\caption{(Color online) Numerical results for the 2-magnon susceptibility as a function of imaginary frequency for $S=1/2$, $S=1$ and $S=2$. The solid curves are fits of Eq. (\ref{eq:fitmodel}).}
\label{fig:2MagnonImaginary}
\end{figure}

\begin{figure}
\includegraphics[width=\linewidth]{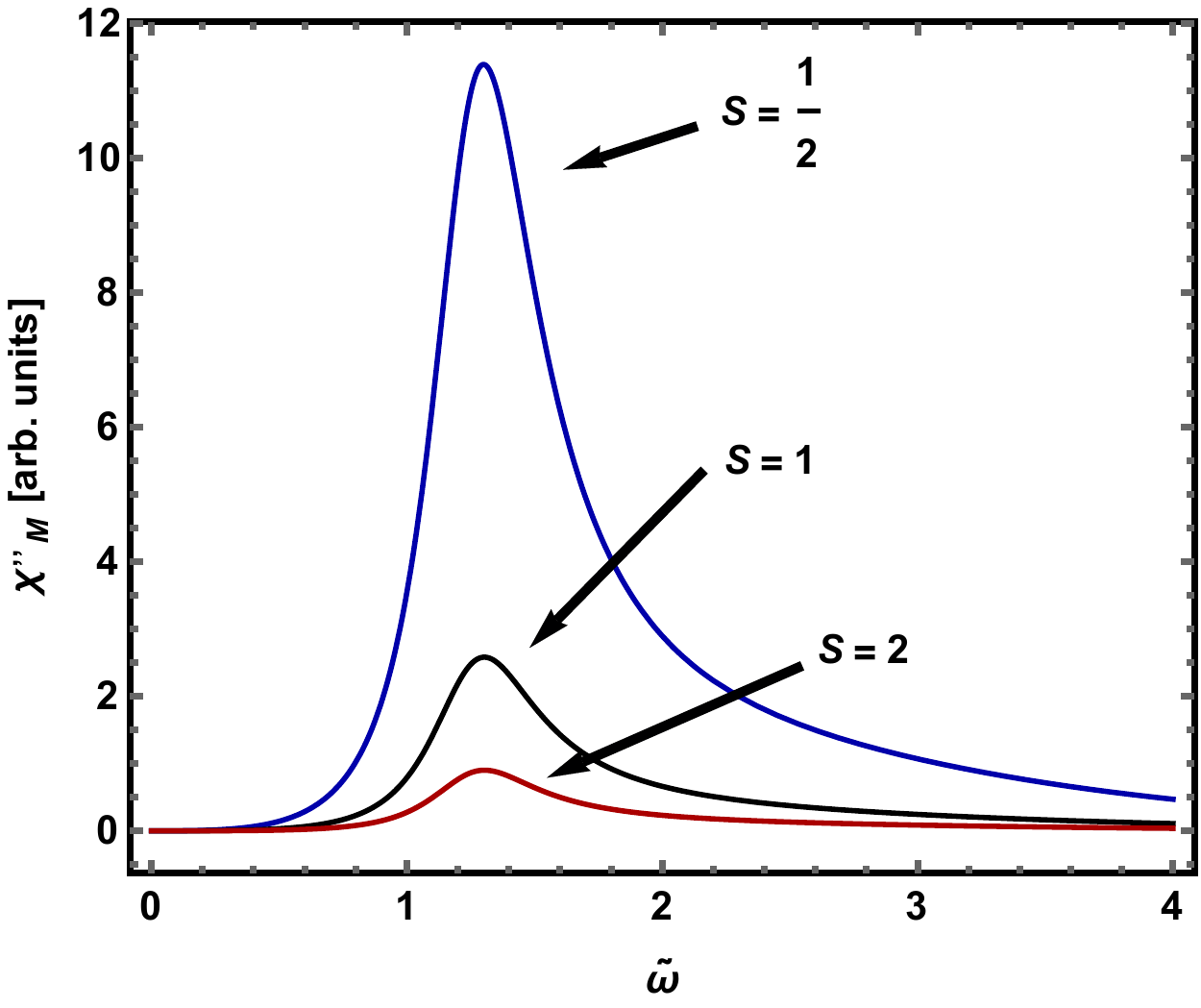}
\caption{(Color online) 2-magnon spectral function $\chi''_{2\text{Magnon}}$ for  $S=1/2$, $S=1$ and $S=2$.}
\label{fig:Spect2Magnon}
\end{figure}

For the numerical solution of the two independent Bethe-Salpeter equations ~\eqref{eq:2MagnonLadderV} and ~\eqref{eq:2HiggsLadderV},
it is convenient to first write them in dimensionless form. This is achieved by introducing  $\tilde{k}=k/\Lambda$, $\tilde{m_0}=m_0/\Lambda$, $\tilde{g}=g\Lambda$, $\tilde{\chi}(\tilde{\omega}) = \chi(\tilde{\omega})/\Lambda$ and $\tilde{\Gamma}(\tilde{\mathbf{k}},\tilde{\mathbf{k}}', \tilde{\omega})=\Lambda^{-2}\Gamma(\tilde{\mathbf{k}}, \tilde{\mathbf{k}}', \tilde{\omega})$.The dimensionless frequency $\tilde{\omega}=\omega/(c_s\Lambda)$ is measured in units of $c_s\Lambda=\pi\hbar J \sim S$. Note that the dimensionless coupling $\tilde{g}=g\Lambda$ is related to the spin quantum number $S$ by $\tilde{g}=2\pi/S$, which is far from small in the relevant case $S=1/2$ discussed below. In practice, the relation $\tilde{g}=2\pi/S$, which arises from a semi-classical expansion valid for $S\gg 1$, does not enter our results for the Raman spectra in quantitative terms, except for determining the relative weight and frequency scales for different $S$ in Figs.~\ref{fig:2MagnonImaginary} and~\ref{fig:Spect2Magnon}. For a given value of $S$, our expansion to leading order in $g$ is therefore valid provided $\tilde{g}\ll\tilde{g}_c$, i.e. the antiferromagnet is far away from a complete destruction of N\'eel order by quantum fluctuations~\cite{PRB.39.2344}. 


The explicit numerical solution of the Bethe-Salpeter equation employs the standard approach to integral equations, namely discretizing the integrals and solving the resulting system of linear equations. For the dimensionless Brillouin-zone $\left[-1,1\right]^2$, we use $20\times20$ points. 
Since an increase up to a grid size of $30\times30$ gave only very small corrections, we can assume that the results have converged to the
continuum limit. While the frequency integration has no intrinsic microscopic cut-off, it is restricted in practice to values below $4c_s\Lambda$. Contributions beyond that do not affect our results. We have varied the number of sampling points for the frequency integral between $20$ and $40$.  Similar to the situation for the momentum integration, 
 the grid size had no influence on the results.   

In FIG.~\ref{fig:2MagnonImaginary}, the two-magnon susceptibility $\tilde{\chi}_{\text{M}}(\tilde{\omega})$ obtained from the numerical solution of 
the Bethe-Salpeter equation is shown for three different values of the dimensionless coupling $\tilde{g}$, corresponding to $S=1/2, 1, 2$ and for an 
intermediate value of the Higgs mass $\tilde{m}_0=0.5$. To obtain the associated spectral function, which determines the Raman spectrum
in real frequencies, an analytic continuation is required. Since our numerical data contain no noise, we use an Ansatz to fit the data which can 
then be continued in analytical form.  Specifically, as a fit model, we use 
\begin{equation}
\chi(\tilde{\omega})=\frac{a_1+a_2q^2}{1+a_3q^2+a_4q^3+a_5q^4} \text{ ,   }q=\sqrt{\tilde{\omega}^2}
\label{eq:fitmodel}
\end{equation}
This Ansatz is motivated by our leading order results for the two-magnon susceptibility, which obeys $\chi_{\text{M}}\sim\tilde{\omega}^{-2}$ for large $\tilde{\omega}$ and $\tilde{\chi}''_{\text{M}}\sim \tilde{\omega}^3$ for small $\tilde{\omega}$.
The fitted curves are also shown in FIG.~\ref{fig:2MagnonImaginary}.  Apparently they reproduce the numerical results very well with a variance which is only $1.4\times10^{-6}$.

Based on the explicit form of the spectral function Eq.~\eqref{eq:fitmodel} in terms of the imaginary frequency variable $q$, 
the analytic continuation $q\rightarrow -i\tilde{\omega}$ is easily done. The resulting two-magnon spectral functions $\tilde{\chi}_{\text{M}}''(\tilde{\omega})$ for $S=\frac{1}{2}$, $1$, $2$ are shown in FIG.~\ref{fig:Spect2Magnon}. For all values of $S$, a clear two-magnon peak appears. 
In the most relevant case $S=1/2$, its maximum is
at $\tilde{\omega}_{\rm max}\simeq 1.29$.  Expressed in terms of the exchange coupling $J$, this 
translates into a frequency $\omega_{\rm max}\simeq 2.44\, J$, somewhat smaller 
than the result $\omega_{\rm max}\simeq 2.76\, J$ obtained from an interacting spin-wave theory~\cite{J.Phys.C.2.2012}. As
will be shown below, the experimental spectra are consistent with our present result. Indeed the observed peak in the Raman spectra 
are close to $2.44\, J$ with values of the exchange coupling which agree very well with those determined
independently from neutron scattering. 
Taking into account the different frequency scales for different values of $S$,
the peak is shifted towards lower frequencies for smaller spin. This behavior has previously also been observed by Canali and Girvin\cite{PRB.45.7127}. 
A quite significant difference between their results and ours, however, is the pronounced asymmetry of the two-magnon peak.  As will be shown
in the following section, this asymmetry is essential for achieving a quantitative agreement with experiment.\\

In order to see to which extent the value of the Higgs mass $\tilde{m}_0$ influences our results for the two-magnon response,
we have varied this parameter in the range $0.1<\tilde{m}_0<0.9$.  Surprisingly, we find no detectable change in the two-magnon line shape. 
The physical reason for the fact that the two-magnon spectral function is essentially independent of the Higgs mass despite the fact that the magnon interaction is mediated by the amplitude mode, is the following: The inverse Higgs mass is the range of the interaction in position space while the interaction strength is controlled by the coupling $g$. As both magnons, involved in the interaction process, are created by a single photon, they are initially on neighboring lattice sites. For $\tilde{m}_0<1$ the range of the Higgs-mediated interaction is larger than one lattice spacing, hence we do not expect a sensitive dependence of the 2-magnon peak on the Higgs mass.

\begin{figure}
\includegraphics[width=\linewidth, height=200pt]{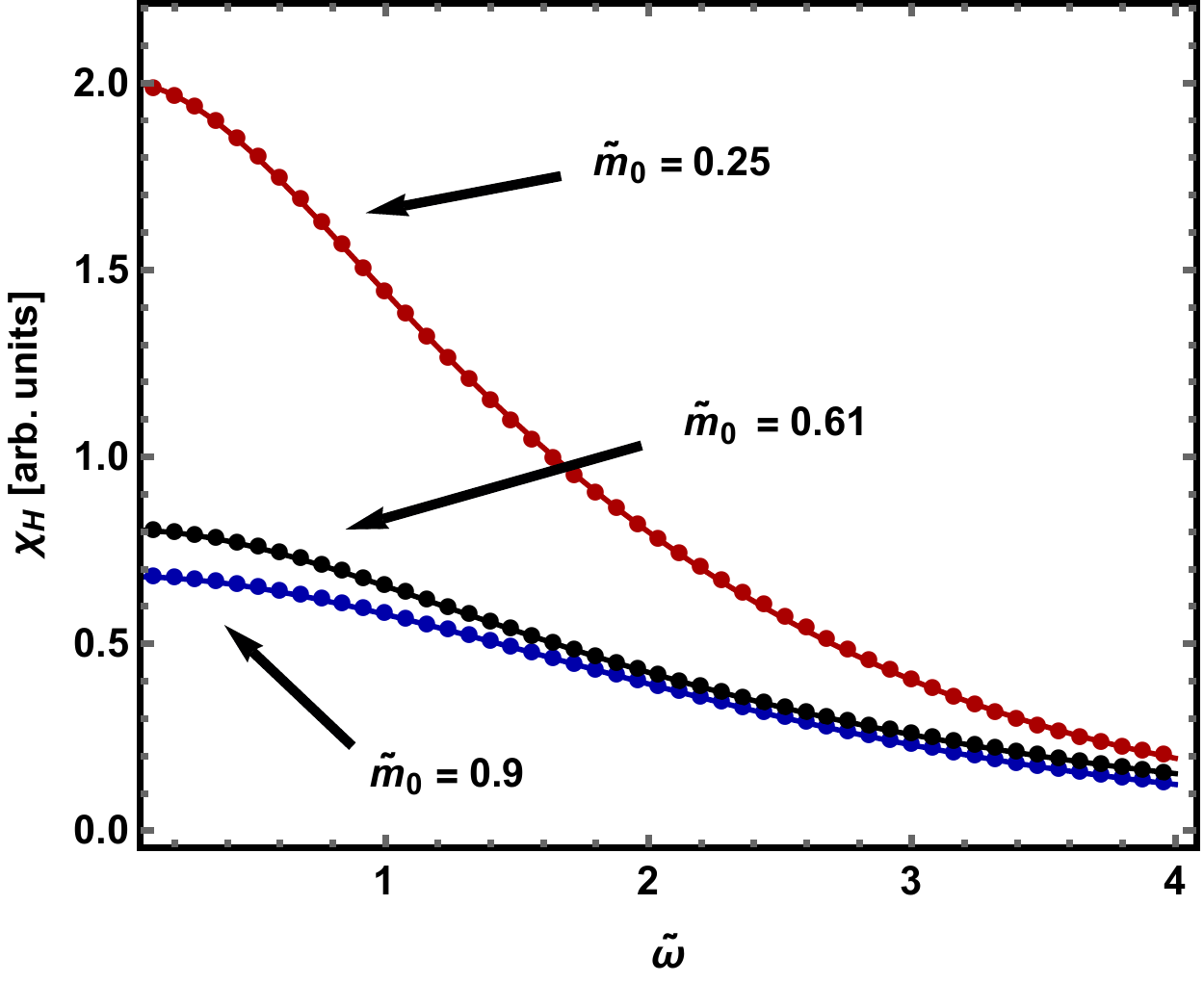}
\caption{(Color online) Numerical results for the 2-Higgs susceptibility as a function of imaginary frequency for $\tilde{m}_0=0.9$, $\tilde{m}_0=0.61$ and $\tilde{m}_0=0.25$. The solid curves are fits of eq.~\eqref{eq:fitmodel}.}
\label{fig:2HiggsImaginary}
\end{figure}

\begin{figure}
\includegraphics[width=\linewidth]{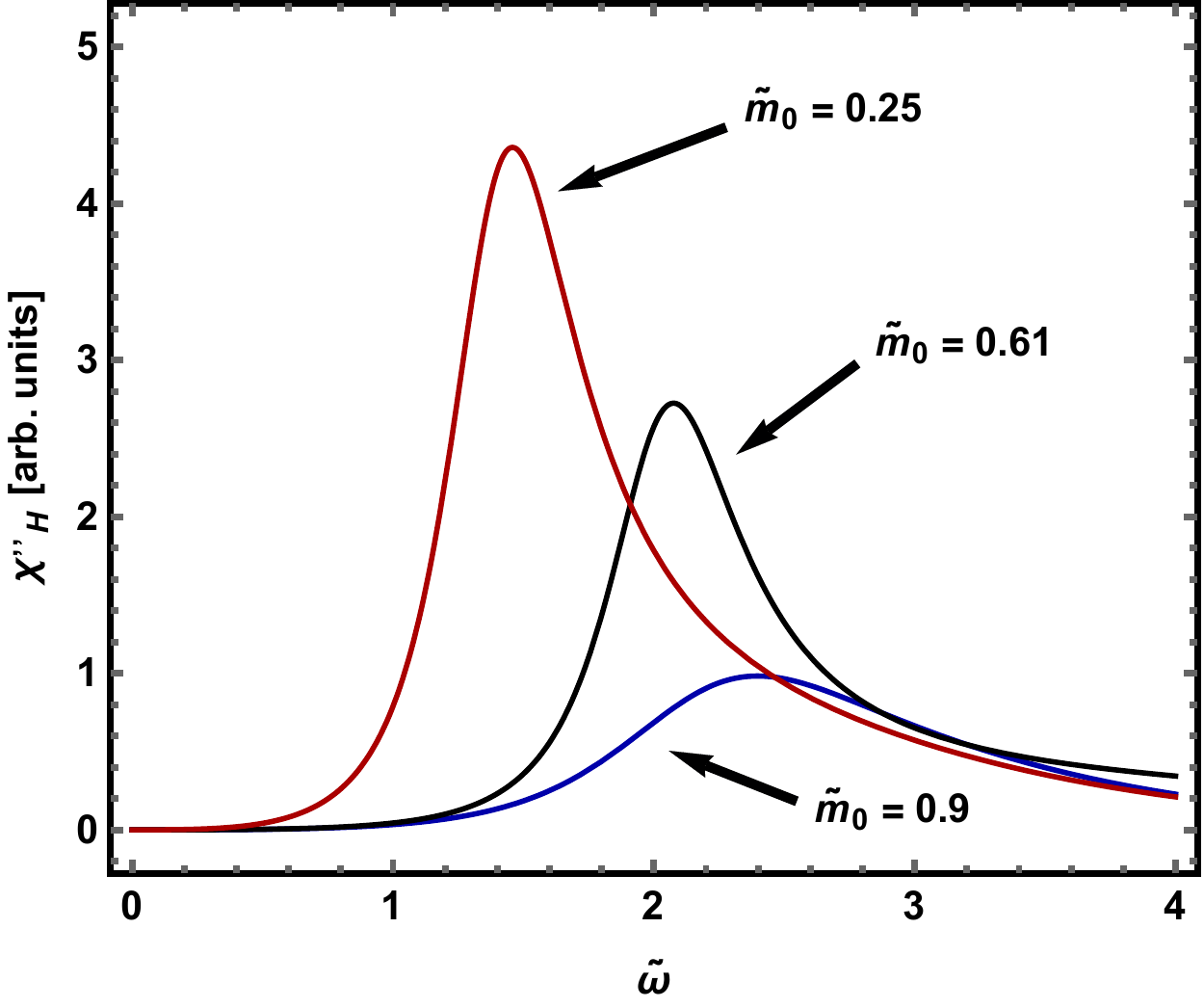}
\caption{(Color online) 2-Higgs spectral function $\chi''_{\text{H}}$ for $S=1/2$ and three different values of the Higgs-mass, $\tilde{m}_0=0.9$, $\tilde{m}_0=0.61$ and $\tilde{m}_0=0.25$.}
\label{fig:Spect2Higgs}
\end{figure}

As a second step, we turn to the results for the two-Higgs susceptibility where a similar dependence on the spin quantum number $S$ appears as 
for the two-magnon response.  We therefore only show the results for $S=1/2$ which is relevant for the comparison with experiments below. 
In FIG.~\ref{fig:2HiggsImaginary} the imaginary time data for the two-Higgs susceptibility is shown for different Higgs masses. First of all one notices, that the absolute values of the two-Higgs susceptibility are smaller than the ones for the two-magnon susceptibility. As a result, the two-magnon peak will 
always dominate the Raman spectrum. From the leading order analysis, we again expect a $\tilde{\omega}^{-2}$-behavior for large frequencies. 
An Ansatz of the form used in Eq.~\eqref{eq:fitmodel} again reproduces the numerical data rather well, with a variance $1.8\times10^{-5}$. 
The resulting conribution $\tilde{\chi}''_{\text{H}}(\tilde{\omega})$ to the Raman spectrum in real frequencies is shown in FIG.~\ref{fig:Spect2Higgs}. 
As expected, the spectra are shifted towards higher frequencies with increasing values of the bare Higgs mass $\tilde{m}_0$.
In particular, the spectral weight is rather small below $2\tilde{m}_0$, reminiscent of the threshold behavior in leading order. 

An exact result which determines the relative weight with which magnons or the Higgs mode appear 
in the Raman spectrum can be obtained by 
considering the ratio of the first moments of the two-magnon and two-Higgs susceptibilities $\chi''_\text{M}$ and $\chi''_\text{H}$. These moments can be expressed in terms of the exact progators $G_{\pi\pi}(k)$ and $G_{\sigma\sigma}(k)$
in a way which is analogous to the standard f-sum rule for the dynamic structure factor. In explicit form, the sum rules are

\begin{align}
&M_\text{M}^{(1)}=\int\limits_{0}^{\infty}\frac{d\omega}{2\pi}~ \omega \chi_\text{M}''(\omega)=2g \int \frac{d^3k}{(2\pi)^3} ~\gamma\left(\mathbf{k}\right) ^2 G_{\pi\pi}(k)\label{eq:MagnonSum}\\
&M_\text{H}^{(1)}=\int\limits_{0}^{\infty}\frac{d\omega}{2\pi} ~\omega \chi_\text{H}''(\omega)=2g \int \frac{d^3k}{(2\pi)^3} ~\gamma\left(\mathbf{k}\right) ^2 G_{\sigma\sigma}(k)\label{eq:HiggsSum}
\end{align}

\begin{figure}
\includegraphics[width=\linewidth]{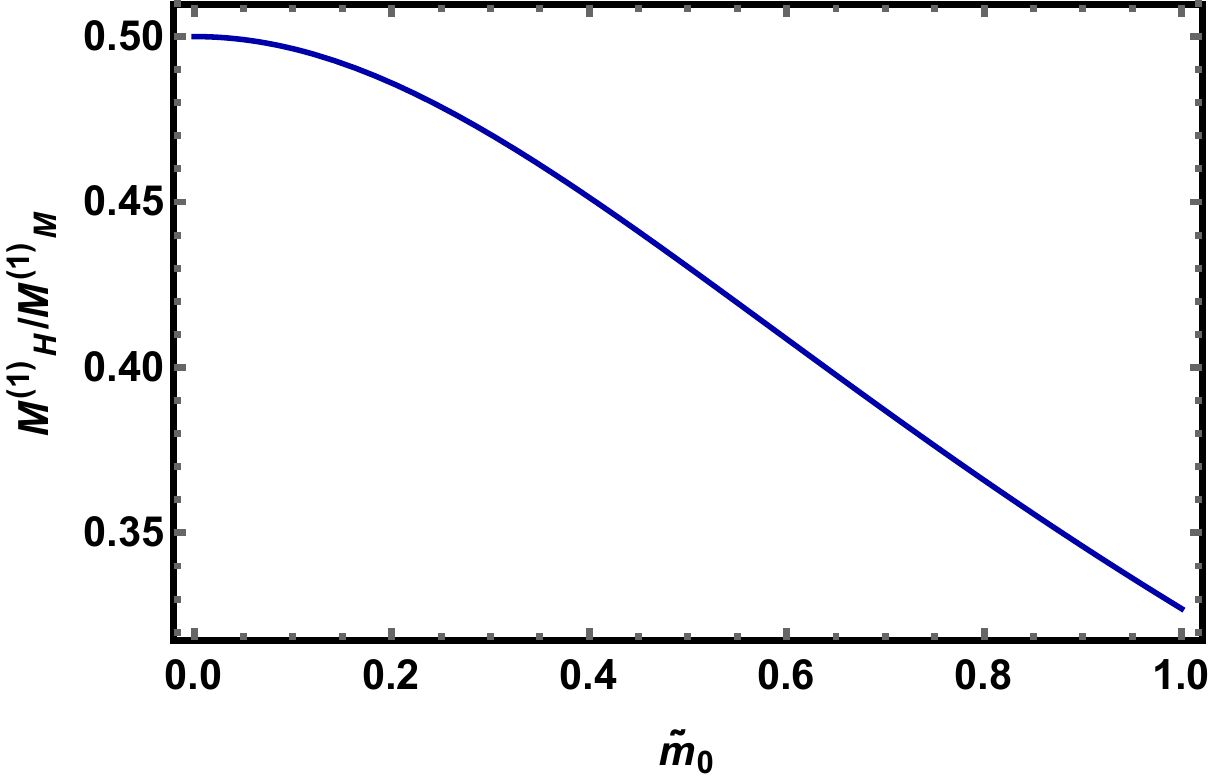}
\caption{(Color online) Ratio of the first moments, $M_\text{H}^{(1)}$ and $M_\text{M}^{(1)}$, of the two-Higgs and two-magnon susceptibility, respectively versus the Higgs mass  $\tilde{m}_0=m_0/\Lambda$ in units of the momentum cut-off at leading order.}
\label{fig:FirstMoment}
\end{figure}

Using the propagators to zeroth order, the ratio between the Higgs and the magnon part of the Raman response 
is controlled by the value of the Higgs mass $m_0$.
As shown in FIG.~\ref{fig:FirstMoment}, the weight of the spectrum coming from the amplitude mode decreases continuously with
increasing mass as expected on intuitive grounds.  

\begin{figure}
\includegraphics[width=\linewidth]{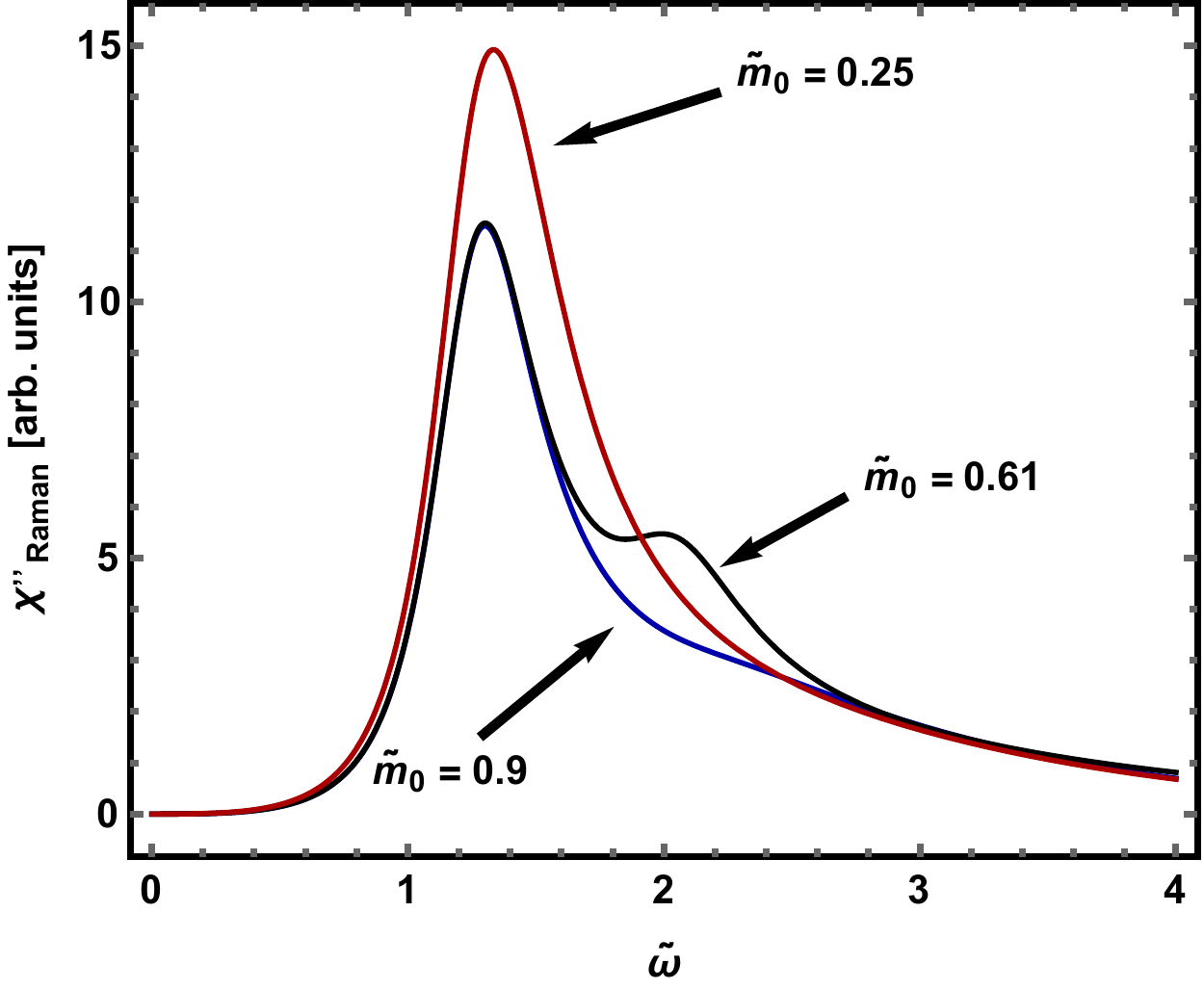}
\caption{(Color online) Total Raman-spectrum, $\tilde{\chi}''_\text{Raman}(\tilde{\omega})=\tilde{\chi}''_{\text{M}}(\tilde{\omega})+\tilde{\chi}''_{\text{H}}(\omega)$, for three different values of the Higgs-mass $\tilde{m}_0$ and $S=1/2$. $\tilde{m}_0=0.25$ is an example for a light Higgs, $\tilde{m}_0=0.61$ for a intermediate Higgs and $\tilde{m}_0=0.9$ for a heavy Higgs.}
\label{fig:ChiSpectRaman}
\end{figure}

Finally, we consider the total Raman spectral function\linebreak $\tilde{\chi}''_\text{Raman}=\tilde{\chi}''_{\text{M}}+\tilde{\chi}''_{\text{H}}$ without the interference term, which is negligible in the regime $\tilde{g}\ll\tilde{g}_c$ studied here. 
We again concentrate on $S=1/2$ and show the Raman spectra for three different values of $\tilde{m}_0$ in FIG.~\ref{fig:ChiSpectRaman}.
These values correspond to a light, an intermediate and a heavy Higgs mass. For both a light and a heavy Higgs with
$\tilde{m}_0\lesssim 0.35$ or  $\tilde{m}_0\gtrsim 0.8$, the full spectrum is dominated by the response associated with magnons.
The contribution from $\tilde{\chi}''_{\text{H}}$ only appears as an additional spectral weight at frequencies above the two magnon peak,
while no distinct peak is associated with an excitation of the amplitude mode.  It is only in the intermediate regime
$0.5\lesssim \tilde{m}_0\lesssim 0.7$ where a shoulder appears above the two-magnon peak as a remnant of the sharp onset of a Higgs-mode 
in the leading order calculation of Eq.~\eqref{eq:2HiggsLO}. As will be discussed below, such a feature is present in the experimental Raman 
spectrum of La$_2$CuO$_4$.

\subsection{Comparison with experiment}
\label{sec:Comparison with experiment}

In the following, we compare our theoretical results for the Raman spectral function $\chi''_\text{Raman}$ with experimental data for the undoped cuprate materials YBa$_2$Cu$_3$O$_6$ (Y-123)~\cite{PRB.78.020511} and\linebreak La$_2$CuO$_4$  (LCO) \cite{EPJST.188.131}. 
Regarding the dominant two magnon response, the relevant energy scale is set by the characteristic frequency 
$c_s\Lambda$ of the underlying linear $O(3)$ $\sigma$-model, which is related to the 
exchange coupling by $\hbar J=\pi^{-1}c_s\Lambda$
\footnote{In our units $\left[J\right]=(\mathrm{Joule})^{-1}s^{-2}$. It is common to give $J$ in units of energy, which we will adopt in the following, i.e. effectively we calculate $\hbar^2 J$.}  For undoped curates, the appropriate values for $J$ are known from inelastic neutron scattering. The agreement between the 
optimal values for $J$ obtained from our fit to the Raman data and the n-scattering results is therefore a crucial consistency test for both the model and 
our approximations involved in calculating the Raman spectrum. Since neither experiments nor theory can reliably
determine the Raman intensity in physical units of counts per second and per frequency interval, 
the results are plotted in arbitrary units.  Correspondingly, the experimental data are fitted with a function of the form 
$A \tilde{\chi}''\left(\omega/(c_s\Lambda)\right)$, where an overall prefactor $A$ remains undetermined. It is important to note however, that this arbitrary overall factor does not affect the relative weight to the Raman spectrum which comes from the magnons or the Higgs mode.  
Apart from the dimensionless Higgs mass $\tilde{m}_0$ this leaves the frequency scale $c_s\Lambda=\pi\hbar J$ as the 
only physical parameter which the spectra depend on. An important and unexpected result of the comparison between theory 
and experiment below is that without the Higgs mode contribution the slow decay of the spectrum above 
the two magnon peak cannot be described properly. The amplitude mode of 2D quantum antiferromagnets is therefore
directly visible in the Raman spectrum despite the fact that generically it does not show up as a separate peak.  It is only
in the particular case of LCO, where the Higgs contribution gives rise to an additional shoulder in the tail above the 
two magnon peak which is likely due to an intermediate mass Higgs. \\  

\begin{figure}
\includegraphics[width=\linewidth]{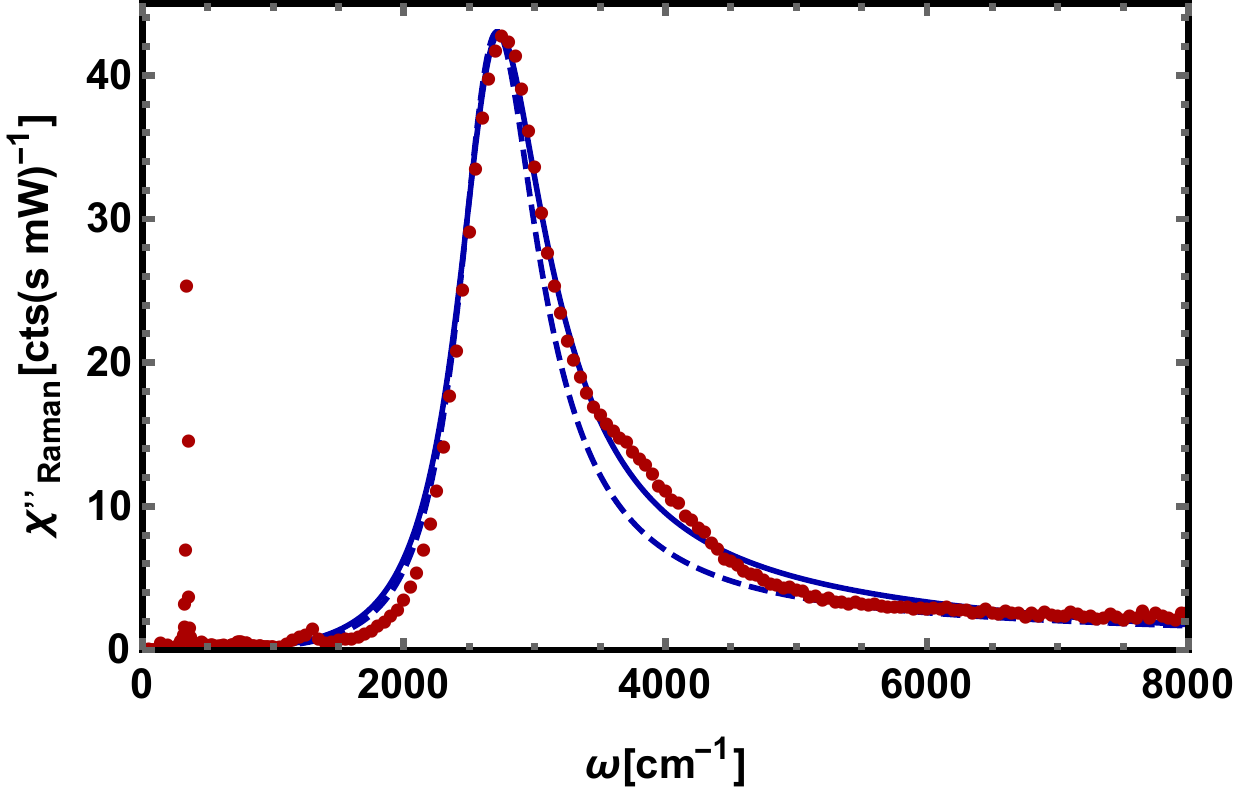}
\caption{(Color online) Experimental data for the Raman spectrum of Y-123 together with a fit of the theoretical model with Higgs mass $\tilde{m}_0=0.25$ (solid line) and without $\chi_\text{H}''$ (dashed line).}
\label{fig:Y123Fit}
\end{figure}

The experimental data for the bilayer compound Y-123 are shown in FIG.~\ref{fig:Y123Fit} together with theoretical fits of the full Raman 
susceptibility. Apparently, the quite non-trivial form of the dominant two-magnon peak is reproduced quite well by our theory with 
an exchange coupling $J=126\mathrm{meV}$ which is rather close to the value $120\mathrm{meV}$ obtained from inelastic 
neutron scattering~\cite{PRB.53.R14741}.  The optimal value of the dimensionless Higgs mass for Y-123 is $\tilde{m}_0=0.25$.
This is significantly smaller than the value $\tilde{m}_0\simeq 0.6$ which is found below for LCO.
A possible origin for the small value of the Higgs mass, which entails large fluctuations of the 
magnitude of the N\'eel  order, may be that Y-123 is a bi-layer system  with an antiferromagnetic interlayer exchange coupling 
$J_\perp/J\approx 0.1$~\cite{PRB.53.R11930, PRB.53.R14741}.  Such a system undergoes a quantum phase transition from a phase of 
two weakly coupled 2D antiferromagnetic layers to a gapped singlet phase with increasing 
$J_\perp$~\cite{PRB.52.3521, PRL.72.2777, EPL.42.559}. 
While Y-123 is clearly in the phase where the two layers possess antiferromagnetic order, the interlayer coupling is expected to 
increase quantum fluctuations in the N\'eel order compared to the single-layer compound LCO and hence leads to 
a smaller Higgs mass.  The presence of 
an amplitude mode directly shows up in the Raman spectrum via the slow decay above the two magnon peak. Indeed,   
the dashed line FIG.~\ref{fig:Y123Fit}, which is an optimal fit to the spectrum without the Higgs, clearly misses a substantial part 
of the spectral weight in the range between $3000\mathrm{cm}^{-1}$ and $5000\mathrm{cm}^{-1}$. 
A feature which is not poperly described by our model is the slightly non-monotonic decay in this frequency range. Such a 
nontrivial structure may in principle be caused by a triple resonance discussed by Morr and Chubukov~\cite{PRB.56.9134}. 
This resonance appears in situations where the incoming photon energy $\hbar\omega_i$ is above the charge transfer gap $2\Delta$. 
The pure spin-photon interaction of the Fleury-Loudon Hamiltonian then needs to be extended by taking into account
the generation of intermediate particle-hole pairs. While the experiments indeed were in a range with $\hbar\omega_i>2\Delta\simeq 2\mathrm{eV}$,
they do not show a change of the additional feature with increasing values of the incident photon energy, as expected for a triple resonance~\cite{PRB.56.9134}.  An explanation of the slightly non-monotonic decay of the spectrum in terms of a triple resonance therefore seems unlikely.       

\begin{figure}
\includegraphics[width=\linewidth]{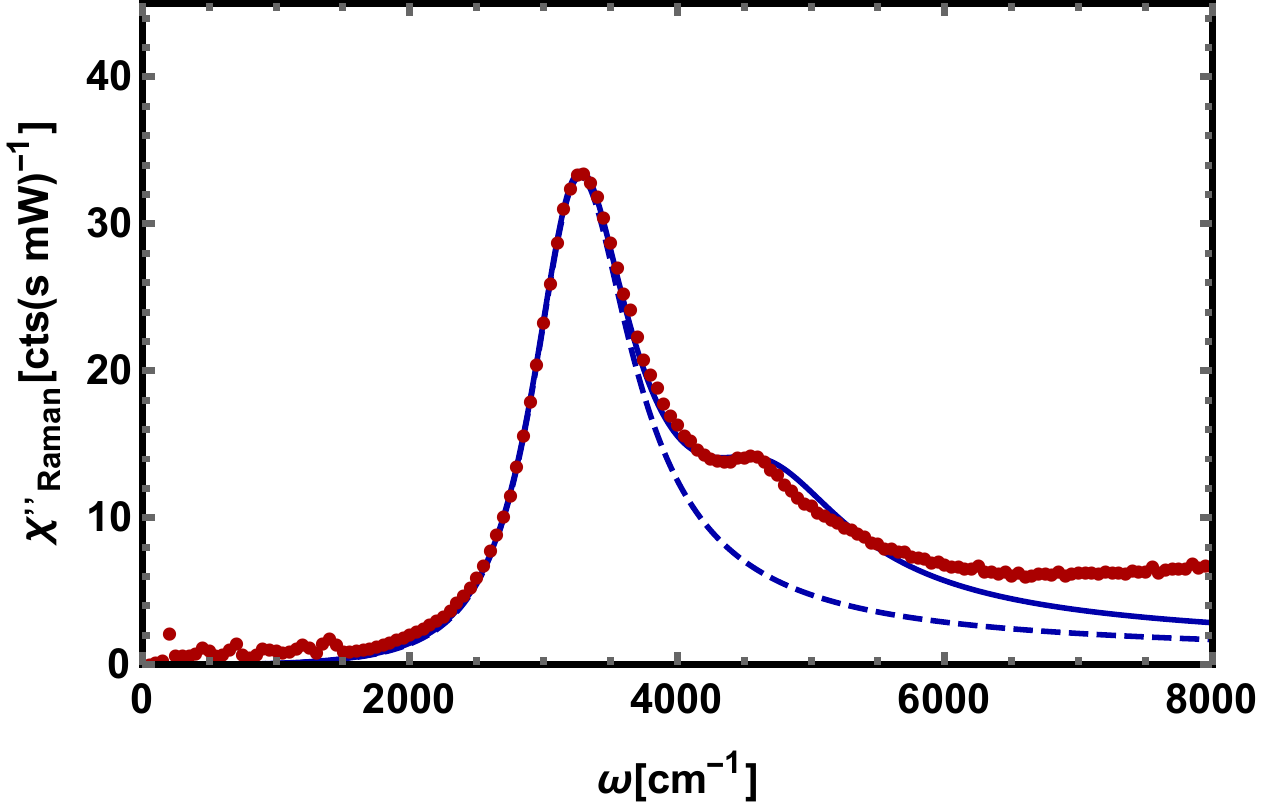}
\caption{(Color online) Experimental data for the Raman spectrum of LCO together with a fit of the theoretical model with Higgs mass $\tilde{m}_0=0.595$ (solid line) and without $\chi_\text{H}''$ (dashed line)}
\label{fig:LCOFit} 
\end{figure}

For the single-layer compound LCO the data together with the theoretical fit are shown in FIG.~\ref{fig:LCOFit}. 
Again, the two-magnon peak is reproduced very well 
with a value $J=149\mathrm{meV}$ for the exchange coupling which is in excellent agreement with the INS result $143\mathrm{meV}$~\cite{PRL.105.247001}. Contrary to the case of Y-123 and our theoretical result, the Raman spectrum of LCO exhibits a slight increase with frequency in the spectral range above $6000 \mathrm{cm}^{-1}$. This deviation is most likely due to the luminescence of defects in the LCO sample,
an effect which is absent in the Y-123  spectra shown in FIG.~\ref{fig:Y123Fit}  because of a higher sample quality~\cite{HacklPrivate}.
Concerning the pronounced shoulder above the two-magnon peak at about $4500\mathrm{cm}^{-1}$, it turns out that this additional feature is reproduced quite well by choosing an intermediate value $\tilde{m}_0=0.595$ of the Higgs mass.   
The overall very good agreement between theory and experiment suggests, that the additional feature seen in the Raman spectrum of LCO is indeed associated with an amplitude mode of the underlying N\'eel state. 
What remains to be understood, however, is the fact that the pronounced feature the $B_{1g}$ mode near $4500\mathrm{cm}^{-1}$ in LCO also appears in the experimental data for the $A_{2g}$ mode, where no two-magnon response is visible. 
By contrast, there is essentially no Raman signal at any frequency in the $A_{2g}$ mode of Y-123~\cite{EPJST.188.131}.

\section{Conclusion and Open Problems}
\label{sec:Conclusion and Open Problems}

In summary, we have shown that the simple linear $O(3)$ $\sigma$-model is able to explain quantitatively the 
major features of the Raman spectrum of the N\'eel state of undoped cuprates. In particular,
we have found that the detailed form of the dominant two-magnon peak
is a result of magnon-magnon interactions which are mediated by the 
Higgs mode. The  relevant energy scale is fixed by the exchange coupling of the 
underlying Heisenberg model, with the peak located at  $\omega\simeq 2.44\, J$.
Apart from mediating interactions between the magnons, the Higgs mode also 
shows up more directly in an enhancement of the spectral weight above the 
two magnon peak. It leads to a separate peak only for an intermediate value 
of the Higgs mass, a case which is apparently realized in LCO. The agreement between 
experiment and theory depends quite sensitively on the proper 
inclusion of the Higgs mode contribution. Raman scattering therefore appears to
provide a direct signature for the existence of a Higgs mode in 2D quantum antiferromagnets.\\

While it is
remarkable that a description based on the linear $O(3)$ $\sigma$-model 
which only captures the low energy physics of quantum antiferromagnets 
is able to account for the detailed form of the observed Raman spectrum
in the full range of energies up to $6J$ and beyond, our present study of possible 
signatures of a Higgs mode in 2D quantum antiferromagnets
is certainly only a first step into a more detailed analysis of this problem.
Among the open questions in this context we mention two:  

\begin{enumerate}
\item[a)] from a quite general point of view, what are suitable and experimentally 
accessible correlation functions where the Higgs mode in 2D quantum 
antiferromagnets can be seen in direct form and can also be studied into a regime 
where N\'eel order disappears into some more complex spin liquid state?

\item[b)] regarding the specific model studied above, can one calculate 
the mass parameter $m_0$ from an underlying microscopic model
and - moreover- go beyond the perturbative calculation of the Raman spectrum
which is only reliable deep in the N\'eel phase?
\end{enumerate}

The complexity of the correlation functions which enter 
the Raman spectrum makes both problems  quite challenging. 
In particular, to uniquely identify the presence a Higgs mode from the experimental data 
so far, which include the two cuprate examples discussed above and also
the Iridate compound Sr$_2$IrO$_4$ \cite{Gretarsson}, a better microscopic 
understanding of the Higgs mass parameter is needed, both for single 
and double layer compounds, for which $m_0$ is apparently 
much smaller than in the single layer case. 
Regarding the issue of correlation functions where the Higgs mode 
shows up more directly than in a standard Raman experiment, an 
interesting option appears to be resonant  inelastic 
X-ray scattering (RIXS)~\cite{RevModPhys.83.705}. Of particular interest here
is indirect RIXS, which allows to measure a momentum dependent four-spin correlation associated with the 
operator~\cite{EPL.80.47003}
\begin{equation}
\hat{O}_{\mathbf{q}}=\sum_{\mathbf{k}} J_{\mathbf{k}}\, \hat{\mathbf{S}}_{\mathbf{k}-\mathbf{q}}\cdot\hat{\mathbf{S}}_{-\mathbf{k}}
\label{eq:RIXS}
\end{equation}
In the language of the NL$\sigma$M Eq.~\eqref{eq:RIXS} corresponds to $\hat{O}_\mathbf{q}\sim\int d^2 x ~e^{i\mathbf{q}\cdot\mathbf{x}}\left(\partial_\mu\mathbf{n}\right)^2$. In comparison to Eq.~\eqref{eq:RamanOxy} one notices, that RIXS again couples to the square of field gradients, however in a symmetric form, without the antisymmetric tensor $\sigma^z_{ij}$, which is responsible for the vanishing of the leading diagrams associated with the Higgs mode.
It is likely therefore that the Higgs mode in 2D quantum antiferromagnets
is directly observable in a indirect RIXS experiment, similar to the probe 
of a Higgs mode in the Bose-Hubbard model which is induced essentially by a
modulation of the hopping amplitude~ \cite{nature11255}. By contrast, in 
{\it direct} RIXS one couples to a single spin operator $\hat{O}_\mathbf{q}^\text{(direct)} \sim (\mathbf{e}_i\times\mathbf{e}_f) \cdot \hat{\mathbf{S}}_\mathbf{q}$~\cite{PRL.108.177003}. This is similar to determining the dynamic spin structure factor in inelastic neutron scattering and allows to measure the magnon-dispersion.

\begin{acknowledgements}
We are very grateful for many insightful discussions with 
Assa Auerbach and Daniel Podolsky and for their kind hospitality 
at the Technion during a crucial stage of this work. We are also 
indepted deeply to Rudi Hackl, both for providing the motivation
to start this project in the first place and for long discussions about
the details of the experiments. Finally, it is a pleasure to acknowledge
constructive comments from Tom Devereaux and from Eugene Demler.  

After completion of our work, we have also benefitted a lot from 
discussions with Hlynur Gretarsson, Bernhard Keimer, Ginat Khaliullin and
Mathieu LeTacon on Raman scattering on Iridates, which in fact provide 
further support for the presence of a Higgs mode in 2D quantum antiferromagnets, see~\cite{Gretarsson}.   
\end{acknowledgements}

\bibliography{Higgs-MagnonEPJB.bib}

\begin{thebibliography}{52}
\bibitem [{\citenamefont {Gretarsson}\ and\ \citenamefont
  {etal.}(shed)}]{Gretarsson}%
  \BibitemOpen
  \bibfield  {author} {\bibinfo {author} {\bibfnamefont {H.}~\bibnamefont
  {Gretarsson}}\ and\ \bibinfo {author} {\bibnamefont {etal.}},\ }\href@noop {}
  {\  (\bibinfo {year} {To be published})}\BibitemShut {NoStop}%
\end{thebibliography}

\end{document}